\documentclass[prl,aps,amsmath,amssymb,twocolumn,superscriptaddress]{revtex4-2} 

\usepackage[colorlinks=true,citecolor=blue]{hyperref}
\usepackage{graphicx}
\usepackage{amsmath, amstext, amssymb, amsfonts, amsxtra}
\usepackage{xspace}
\DeclareSymbolFontAlphabet{\amsmathbb}{AMSb} 
\usepackage{xcolor}
\usepackage{braket}

\newcommand{\HL}{H_{\rm L}}
\newcommand{\HR}{H_{\rm R}}

\newcommand{\NL}{N_{\rm L}}
\newcommand{\NR}{N_{\rm R}}
\newcommand{\NLR}{N_{\rm L/R}}

\newcommand{\cur}{\mathcal{I}}

\newcommand{\Ityp}{\cur^{\rm typ}}
\newcommand{\Imic}{\cur^{\rm mic}}
\newcommand{\Iexa}{\cur_{\rm exact}}

\newcommand{\Ieigexa}{\cur^{\rm eig}_{\rm exact}}
\newcommand{\Ieigper}{\cur^{\rm eig}_{\rm perturb}}
\newcommand{\Itypexa}{\cur^{\rm typ}_{\rm exact}}
\newcommand{\Itypper}{\cur^{\rm typ}_{\rm perturb}}
\newcommand{\Imicexa}{\cur^{\rm mic}_{\rm exact}}
\newcommand{\Imicper}{\cur^{\rm mic}_{\rm perturb}}

\newcommand{\tr}{\mathrm{Tr}}

\newcommand{\expe}[1]{\mathrm{e}^{#1}}
\newcommand{\im}{\mathrm{i}}

\newcommand{\Exp}{\mathbb{E}}

\newcommand{\BL}{B^{\rm L}}
\newcommand{\dBL}{\dot{B}^{\rm L}}

\newcommand{\BR}{B^{\rm R}}

\newcommand{\sx}[1]{\sigma^{x}_{#1}}
\newcommand{\sy}[1]{\sigma^{y}_{#1}}
\newcommand{\sz}[1]{\sigma^{z}_{#1}}

\newcommand{\para}[1]{{\em #1}\/.---}

\newcommand{\sutd}{Science, Mathematics and Technology Cluster, Singapore University of Technology and Design, 8 Somapah Road, 487372 Singapore} 
\newcommand{\sutdepd}{EPD Pillar, Singapore University of Technology and Design, 8 Somapah Road, 487372 Singapore} 
\newcommand{\hnu}{Key Laboratory of Low-Dimensional Quantum Structures and Quantum Control of Ministry of Education, Department of Physics and Synergetic Innovation Center for Quantum Effects and Applications, Hunan Normal University, Changsha 410081, China
}

\begin{document}

\title{Typicality of nonequilibrium (quasi-)steady currents}               

\author{Xiansong Xu} 
\affiliation{\sutd}
\author{Chu Guo} 
\affiliation{Henan Key Laboratory of Quantum Information and Cryptography, Zhengzhou,
Henan 450000, China}
\affiliation{\hnu}
\author{Dario Poletti}
\affiliation{\sutd} 
\affiliation{\sutdepd}

\begin{abstract} 
The understanding of the emergence of equilibrium statistical mechanics has progressed significantly thanks to developments from typicality, canonical and dynamical, and from the eigenstate thermalization hypothesis. Here we focus on a nonequilibrium scenario in which two nonintegrable systems prepared in different states are locally and non-extensively coupled to each other. Using both perturbative analysis and numerical exact simulations of up to 28 spin systems, we demonstrate the typical emergence of nonequilibrium (quasi-)steady current for weak coupling between the subsystems. We also identify that these currents originate from a prethermalization mechanism, which is the weak and local breaking of the conservation of the energy for each subsystem.      
\end{abstract}

\maketitle

\para{Introduction} 
Daily experience teaches us that when two large systems prepared in different equilibrium states are put into contact, a long-lasting current, which we will refer to as (quasi-)steady, emerges between them. Furthermore, after this possibly long intermediate time, which depends on the size of the two objects, the overall system relaxes to an equilibrium state. To understand and characterize the emergence of the (quasi-)steady current, generally one considers each system to be described by an equilibrium, canonical or microcanonical, ensemble. However, here we ask ourselves if the systems driving the (quasi-)steady state current can be described by single pure states. 
Advances in ``pure state quantum statistical mechanics'' have shown that equilibration can emerge in isolated quantum systems \cite{GogolinEisert2016, Lloyd1988}.  The seminal papers by Deutsch \cite{Deutsch1991} and Srednicki \cite{Srednicki1994} brought to the eigenstate thermalization hypothesis (ETH), which postulates that, for a nonintegrable system, eigenstates that are close in energy have similar local properties. ETH has been tested in a variety of systems \cite{RigolOlshanii2008} and reviews can be found in Refs. \cite{BorgonoviZelevinsky2016, DAlessioRigol2016, GogolinEisert2016, MoriUeda2018, Deutsch2018, Ueda2020}. Meanwhile, the attention on the foundation of statistical mechanics has led to the development of the notion of typicality \cite{Lloyd1988,  Reimann2007, BartschGemmer2009, GemmerMahler2010}, within canonical formalism \cite{GoldsteinZanghi2006, PopescuWinter2006} and for unitary dynamical processes \cite{BartschGemmer2009,Reimann2018a} with more specific scenarios such as prethermalization \cite{Reimann2018} and perturbation \cite{DabelowReimann2020, DabelowReimann2021}. For example, dynamical typicality states that pure states with the same initial expectation value for some observables will likely have similar expectation values at any later time \cite{BartschGemmer2009,Reimann2018a}. This was used to show that a ``weak'' version of ETH \cite{BiroliLauchli2010, IkedaUeda2013, BeugelingHaque2014, IyodaSagawa2017, YoshizawaSagawa2018} is both necessary and sufficient for the vast majority of states to thermalize \cite{Reimann2018}. 

Within the thermalization dynamics, the system may experience prethermalization \cite{BergesWetterich2004, MoeckelKehrein2008, EcksteinWerner2010, KollarEckstein2011, AokiWerner2014, MallayyaDeRoeck2019a, MallayyaRigol2021}. A key signature of this phenomenon is a separation of timescales during the relaxation, for instance, a fast initial dynamics in which the system relaxes to an intermediate (often called prethermal) state, and a slower relaxation toward the true thermal state. This separation of time scales can be seen in systems with a weakly broken conserved quantity, and it has been explored in ultracold atoms experiments \cite{GringSchmiedmayer2012}. In Ref. \cite{ReimannDabelow2019a} it was shown that prethermalization is typical in the presence of weak coupling. 

Much less is known regarding the emergence of nonequilibrium (quasi-)steady current when coupling two systems in pure states. Here we show the emergence of such typical quasi-steady current based on the notion of dynamical typicality. We highlight that there have been remarkable studies on the emergence of typical dynamics for nonequilibrium systems \cite{MonnaiYuasa2014, MonnaiYuasa2016, EvansRondoni2016}, and also insightful works to extend ETH to open quantum systems \cite{MoudgalyaSondhi2019a, ShiraiMori2020}. However, in our work, we do not assume a priori that a steady current can be reached, and we consider a unified framework for both the emergence of (quasi-)steady currents and thermalization. 

For the two ``baths'' we take two nonintegrable spin chains coupled at one of their edges, and we initialize the baths either in single eigenstates or random pure states taken within an energy shell. For weak coupling between the systems, we show that the resulting (quasi-)steady current is typical in the sense that it converges, when the system size increases, to what would be obtained from initializing the baths in microcanonical states. We also verify that the value of the current converges towards the prediction from the ETH ansatz. Furthermore, we are able to show that the dynamics that leads to the formation of a long-lasting current, and the eventual thermalization of the two coupled chains, can be understood in the framework of prethermalization. In fact the dynamics of each bath, when decoupled, conserves their own energy, while the coupling between the baths breaks this conservation law. For weak coupling between the two chains one thus expects a slow ``prethermalization-like'' dynamics, where the prethermal state actually approaches a nonequilibrium steady state. Finally, we numerically show that the relaxation dynamics is proportional to the square of the coupling between the chains, and inversely proportional to their length, which can be derived within the prethermalization paradigm \cite{MallayyaDeRoeck2019a, MallayyaRigol2021}. We thus expect, in the thermodynamic limit, that the current will exist indefinitely.

\para{Model} 
Two finite-size quantum systems $\HL$ and $\HR$ are treated as the left and right baths, which are coupled via an interaction term $V$, i.e., the total Hamiltonian of the system is $H = \HL + \HR + V$. 
Each bath is chosen to be a nonintegrable spin chain with bond and site Hamiltonian ($h^{\rm b }_{n}$ and $h^{\rm s}_{n}$) given by   
\begin{align}
    h_{n}^{\rm b} = J_{zz} \sz{n}\sz{n+1} + J_{yz} \sy{n}\sz{n+1},\;\;  h_{n}^{\rm s} = h_x \sx{n} + h_z \sz{n}     \label{eq: hamiltonian}
\end{align}
such that $\HL = \sum_{n=1}^{\NL-1} h^{\rm b}_{n} + \sum_{n=1}^{\NL} h^{\rm s}_{n}$, while for $\HR$ the site labelling ranges from $\NL+1$ to $N$. Here $N=\NL+\NR$ is the total number of spins, while $\NL$ and $\NR$ are the length of the left and right bath respectively. 
The interaction term $V$ is given by $V = \gamma \BL \bigotimes \BR = \gamma \sx{\NL} \sx{\NL+1}$ where $\gamma$ is the coupling strength. We consider $J_{yz}=J_{zz}$, $h_x=-1.05J_{zz}$, $h_z=0.5J_{zz}$, for which values each bath is nonintegrable with Gaussian unitary ensemble (GUE) level statistics \cite{AtasRoux2013}. The Hamiltonians $\HL$ and $\HR$, which have no conserved quantities apart from energy, have also been used to study out-of-time-ordered correlators \cite{HuangZhang2019}. In the following we will work in units for which $J_{zz}=\hbar=1$. 

\para{Initial conditions} 
We are interested in the currents generation when the baths are prepared with different local equilibrium properties, e.g., the baths are prepared at different initial energies per spin. More specifically, for each bath, we consider an energy shell 
$\Xi^{\rm L/R} = [E^{\rm L/R} - \Delta^{\rm L/R}/2, E^{\rm L/R} + \Delta^{\rm L/R}/2]$ where $E^{\rm L/R}$ is the average energy and $\Delta^{\rm L/R}$ the width of the shell. 

To this end we consider three different scenarios:

i) {\it eigenstate pairs} -- each bath is prepared at an arbitrary eigenstate corresponding to the shell $\Xi^{\rm L/R}$, i.e.,
\begin{align}
    \Ket{\psi^{\rm eig}} = | i \rangle_{\rm L} \otimes | j \rangle_{\rm R} \label{eq:ic_eigen}   
\end{align}
where $\Ket{i}_{\rm L}$ ($ \Ket{j}_{\rm R}$) is the $i(j)$-th eigenstate of $H_{\rm L}$ ($H_{\rm R}$);

ii) {\it typical-state pairs} -- each bath is a typical superposition of states within the shell, with random complex numbers $c^{\rm L}_i$  and $c^{\rm R}_j$ drawn from the Haar measure, i.e., 
\begin{align}
    \Ket{\psi^{\rm typ}} = \sum_{i, j}   c^{\rm L}_i c^{\rm R}_j \; \Ket{i}_{\rm L} \otimes \Ket{j}_{\rm R}.    
\end{align}

iii) {\it microcanonical ensemble pairs} -- each bath is prepared in a microcanonical state of the energy shell $\Xi^{\rm L/R}$,
\begin{align}
    \rho^{\rm mic} = \rho_{\rm L}^{\rm mic} \otimes \rho_{\rm R}^{\rm mic}
\end{align}
with $\rho_{\rm L/R}^{\rm mic} = 1/d_{\rm L/R}\sum_{E^{\rm L/R}_i } \Ket{i}_{\rm L/R}\Bra{i}_{\rm L/R}$, where $E^{\rm L/R}_i \in \Xi^{\rm L/R}$ and $d_{\rm L/R}$ is the number of states in the shell $\Xi^{\rm L/R}$.   

\para{Energy currents} 
The energy current operator with respect to the left environment is defined as $I^{\rm L} =-{\rm d} \HL/{\rm d}t=-\im \gamma \left[\BL,\HL\right] \otimes \BR =  \dBL \otimes \BR$ where we have defined $\dBL \equiv -\im \gamma \left[\BL,\HL\right]$. The expectation value $\Iexa=\tr\left( I^{\rm L} \rho(t) \right)$ gives the exact current from evolving the initial condition via the full Hamiltonian $H$, and it takes three forms $\Ieigexa$, $\Itypexa$ and $\Imicexa$ depending on whether the initial condition $\rho(0)$ is $\ket{\psi^{\rm eig}}\bra{\psi^{\rm eig}}$, $\ket{\psi^{\rm typ}}\bra{\psi^{\rm typ}}$ or $\rho^{\rm mic}$ \footnote{Note that when discussing the expectation value of the current we drop the superscript ${\rm L}$ because, in the (quasi-)steady regime we are interested in, the current from the left bath equals that to the right bath. More details in \cite{Supp}}. A true bath, by definition, should be infinitely large with a continuous spectrum. In our simulation, we would like to consider systems as large as possible to avoid finite-size effects. This poses a great numerical challenge due to the exponential growth of the space and time complexity. For this purpose, various strategies have been proposed \cite{RigolSingh2006, SteinigewegGemmer2014}. We developed a highly optimized time evolution algorithm that allows to study system sizes up to $N=28$ on a personal computer. Two main numerical techniques are used: i) a Suzuki-Trotter based time evolution algorithm which only requires the storage of a single quantum state; ii) a highly parallelized and cache-friendly implementation of the gate operations (See Ref. \cite{Supp} for details).

Since such exact simulations are numerically demanding, we complement these calculations with a perturbative approach. On top of significantly reducing the time and memory demands of the calculations, the perturbative approach also allows us to gain analytical insights on the typicality of the currents as well as the importance of weak coupling between the baths. It also shares the same spirit of the weak coupling master equations formalism \cite{EspositoGaspard2003b, EspositoGaspard2007, BreuerPetruccione2007, DeVegaAlonso2017, LandiSchaller2021}. In the weak coupling limit $\gamma \rightarrow 0$, the perturbative current expression with respect to different initial conditions can be written as \cite{ZhouZhang2020, Supp}
\begin{align}
    \cur(t)  = & \mathcal{\dot{B}}^{\rm L}(t) \mathcal{B}^{\rm R}(t) - \im \int_{0}^{t} d \tau
     {\Big [ } \overrightarrow{\mathcal{C}}_{\rm L}(t,\tau) \mathcal{C}_{\rm R}(t,\tau) \nonumber\\ 
                                    &-\overleftarrow{\mathcal{C}}_{\rm L}(\tau,t) \mathcal{C}_{\rm R}(\tau,t) {\Big ]}, 
    \label{eq: current}
\end{align}
where $\mathcal{\dot{B}}^{\rm L}(t)=\tr_{\rm L}\left[\rho_{\rm L} \dBL(t) \right]$, $\mathcal{B}^{\rm R}(t)=\tr_{\rm R}\left[\rho_{\rm R} \BR(t) \right]$ and we have defined the two-time correlation functions $\overrightarrow{\mathcal{C}}_{\rm L}(t,\tau) =\tr_{\rm L}\left[\rho_{\rm L} \dBL(t)\BL(\tau)\right]$, $\overleftarrow{\mathcal{C}}_{\rm L}(t,\tau) =\tr_{\rm L}\left[\rho_{\rm L} \BL(\tau)\dBL(t)\right]$, and $ \mathcal{C}_{\rm R}(t,\tau) =\tr_{\rm R}\left[\rho_{\rm R} \BR(t)\BR\right]$. We have also defined $B^{\rm L/R}(t)=e^{\im H_{\rm L/R} t}B^{\rm L/R}e^{-\im H_{\rm L/R} t}$ and the same applies to $\dBL(t)$. We call $\Itypper$, $\Imicper$, $\Ieigper$ as typical, microcanonical, and eigenstate currents respectively depending on the initial conditions used. Note that when equilibrium states or single eigenstates are used, the expression for the current Eq. (\ref{eq: current}) can be significantly simplified and the first term vanishes \cite{Supp}.

\begin{figure}
    \centering
    \includegraphics[width=8.6cm]{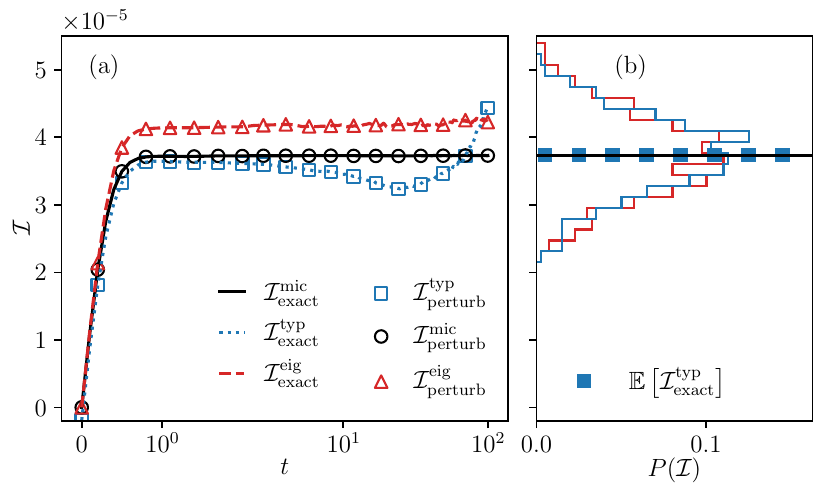}
    \caption{(a) Currents versus time for eigenstate initial conditions (red), microcanonical initial conditions (black), and typical initial conditions (blue). The perturbative results are denoted by markers and the exact currents by lines. (b) Histogram of the current statistics for eigenstate currents and typical currents at $t=10$. Total number of initial states, both eigenstates and typical ones, is 399. The solid black line marks the exact microcanonical current. $N=24$, $\gamma =0.01$, $E_{\rm L}=-E_{\rm R}=5$, $\Delta_{\rm L/R}=0.1$.}
    \label{fig: typicality}
\end{figure}

\para{Typical (quasi-)steady currents} 
We first consider weak coupling $\gamma=0.01$ where we expect the perturbative results to be consistent with the exact ones. In Fig. \ref{fig: typicality} we consider baths each of size $\NL=\NR=12$ and with energy windows $\Xi^{\rm L/R}$ given by $[4.95,5.05]$ and $[-5.05, -4.95]$, respectively. In Fig. \ref{fig: typicality}(a), we depict $\cur$ both for exact and perturbative calculations (respectively for empty symbols and lines), and for the three different initial conditions.  We observe perfect agreement between exact currents and perturbative currents for all the initial conditions considered. For the typical initial conditions, the results present larger oscillations, due to the dynamics of initial coherence for typical states. These results validate the use of perturbative methods for computing currents as all the currents computed are close to each other. In addition, the microcanonical current becomes constant after a time of order $1$ indicating the emergence of a (quasi-)steady current. 

Another question to be answered is whether also the currents $\Itypexa$ or $\Itypper$ can be considered as typical, i.e., the vast majority of the currents using different typical initial conditions show similar dynamics. This means that if we consider an ensemble of typical currents, their average $\Exp\left[ \Ityp \right] \approx \Imic $ with a bounded variance. Perturbatively, one can show this for the (quasi-)steady current in a similar fashion to the study of typicality in isolated quantum systems \cite{Reimann2007, MonnaiYuasa2014}. In fact, the average typical correlations functions $\Exp\left[ \mathcal{C}^{\rm typ} \right]$ is equivalent to the microcanonical ones by considering, for example, 
\begin{align}
    \Exp\left[ \overrightarrow{\mathcal{C}}_{\rm L}^{\rm typ}(t, \tau)\right] =&   -\im \gamma \sum^{D_{\rm L}}_k \sum_{ij}^{d_{\rm L}} \Exp{\left[c_i^* c_j\right]} \expe{\im (\Delta_{ki} t + \Delta_{kj} \tau)} \Delta_{ki} \BL_{ik} \BL_{kj} \nonumber \\
        = & \overrightarrow{\mathcal{C}}_{\rm L}^{\rm mic}(t, \tau)
\end{align}
where $\Exp\left[ c^*_i c_j\right] = \delta_{ij}/d_{\rm L}$, $\Delta_{ki} = E_k - E_i$, $D_{\rm L}$ is the Hilbert space dimension of the left bath, and $\delta_{ij}$ is the Kronecker delta. Since the left and right correlation functions are independent, it is thus straightforward to conclude from Eq. (\ref{eq: current}) that the average typical currents is equal to the microcanonical one, i.e., $\Exp\left[ \Itypper \right] = \Imicper $. This is an indicator of the typicality of (quasi-)steady currents. However, one should also consider the variance of the typical currents. To this end, we perform numerical computations of the exact currents $\Itypexa$ and  $\Ieigexa$ at a time $t=10$, and provide a histogram of these values in Fig. \ref{fig: typicality}(b). Already for bath sizes $\NLR=12$, we observe a clear peak near the prediction from microcanonical initial states (continuous black line) which corresponds to the average of the typical states too (solid blue square). 

\begin{figure}
    \centering
    \includegraphics[width=8.6cm]{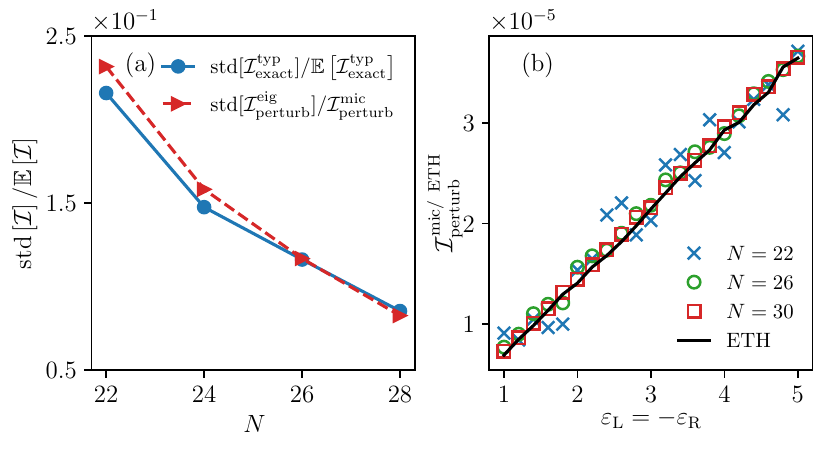}
    \caption{(a) Normalized standard deviation versus the total system size $N$ for exact typical currents (blue) and perturbative eigenstate currents (orange) at time $t=10$. The number of typical initial conditions is 500. (b) Perturbative microcanonical currents versus energy per spin $\varepsilon_{\rm L} = - \varepsilon_{\rm R}$ and ETH predicted current (solid black line) for $N=2000$. $\gamma =0.01$. }     
    \label{fig: ETH}
\end{figure}

We then analyze the dependence of the variance with the size of the baths. Relying on the extensive character of $\HL$ and $\HR$, we linearly shift the energy shell for the initial conditions as the baths' size increases, i.e., $\Xi^{\rm L/R}$ are such that $E^{\rm L/R} = \NLR \varepsilon_{\rm L/R}$ and $\Delta^{\rm L/R}= \NLR \delta_{\rm L/R}$, where $\varepsilon_{\rm L/R}$ is the energy per spin in each bath and $\delta_{\rm L/R}$ is the shell width per spin chosen to be $0.1/12$. In Fig. \ref{fig: ETH} (a), we show the standard deviation for different realizations of typical currents and eigenstate currents ${\rm std}[\mathcal{I}]$ (rescaled over the value of the current $\Exp\left[\mathcal{I}\right]$) versus system size $N$. Such decaying behavior serves as numerical evidence of the typicality of nonequilibrium currents \footnote{The evolution of the currents from the corresponding different typical initial conditions are shown in \cite{Supp} for time up to $t=50$}. Fig. \ref{fig: ETH}(b), which depicts the value of $\Imicper$ (to which the typical current converges) for different energy per spins, also shows a clear convergence of the current with the system size. In fact, larger systems sizes, each denoted by different markers, result in smaller oscillations and the current versus energy curve converges towards a smooth line. To have a deeper understanding of the emergence of this smooth curve, we compare the numerical perturbative results with predictions from ETH. The ETH ansatz, which each bath follows \cite{Supp}, states that for a local observable $O$, its matrix elements in the energy basis are  \cite{Srednicki1994, Srednicki1996}  
\begin{align} 
    O_{ij} &= \Braket{O}(E) \delta_{ij} + \expe{-\frac{S(E)}{2}} f(E,\omega) R_{ij} \label{eq:ETH} 
\end{align}
where $\omega=E_j-E_i$ and $E=(E_i+E_j)/2$, $E_i$ are energy eigenvalues, $S(E)$ is the entropy given by $\expe{S(E)} = E \sum_i \delta(E - E_i)$, $f(E,\omega)$ is a mildly varying function of $E$ and $\omega$ and $R_{ij}$ is a matrix with normal distributed random elements. 
We thus estimate the smooth function $f(E,\omega)$ for each bath from a system of size $N=24$ and evaluate the correlations stemming from Eq. (\ref{eq: current}), $\overrightarrow{\mathcal{C}}_{\rm L}^{\rm ETH}$, $ \overleftarrow{\mathcal{C}}_{\rm L}^{\rm ETH}$ and $\mathcal{C}_{\rm R}^{\rm ETH}$ \cite{DAlessioRigol2016,LuitzBarLev2016,Supp}, from which we can determine the current expected from a large system. In Fig. \ref{fig: ETH}(b) we thus show with a continuous black line the value from this ETH-based approach for a system with $N=2000$. Indeed, the microcanonical currents converge towards ETH predictions showing that the emergence of a smooth dependence of the current versus energy can be deduced from ETH.     

\begin{figure}
    \centering
    \includegraphics[width=8.6cm]{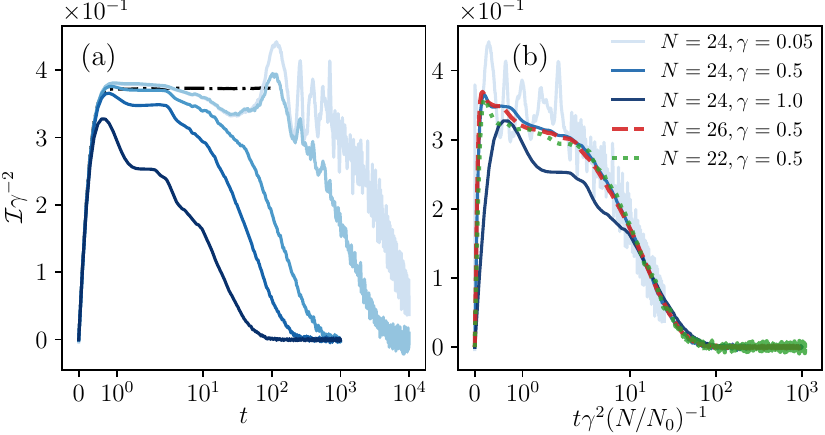}
    \caption{(a) Typical exact currents $\Itypexa$ versus time $t$ for different coupling strengths $\gamma=0.05, \;0.1, \;0.3, \;0.5$, and $\gamma=1.0$ denoted by the increasingly darker blue solid lines. The typical initial states are chosen to be the same with the system size $N=24$. The reference dynamics for perturbative microcanonical current is represented by the black dashed-dotted line. (b) Typical exact currents $\Itypexa$ versus scaled time $t\gamma^2 /(N/N_0)^{-1}$ for different system size and coupling strength with $N_0=24$.}
    \label{fig: coupling}   
\end{figure}

\para{Prethermalization and thermalization} 
To address how the system goes from the presence of a (quasi-)steady current to the absence of current, we consider larger magnitudes of the coupling strength $\gamma$ and we perform exact simulations. We stress here that perturbative calculations do not allow one to observe thermalization because, implicitly, they assume that one takes the thermodynamic limit first, and then the infinite-time limit.
In Fig. \ref{fig: coupling}(a) we depict the exact typical current versus time, for a single random typical initial condition, divided by $\gamma^2$ (because of the dependence on $\gamma$ of the perturbative current) for different values of the coupling strength with $N=24$. In particular we have $\gamma = 0.05,\;0.1,\;0.3,\;0.5$ and $1.0$ for darker shades of blue. As a reference, we add the microcanonical perturbative current $\cur^{\rm mic}_{\rm perturb}$ as a black dot-dashed line. The typical current of a single random realization and for small $\gamma$ follows closely the perturbative current before deviating from it. As $\gamma$ increases, the typical current deviates earlier, and more markedly from the perturbative value. 
In Fig. \ref{fig: coupling}(b) we consider different coupling strengths $\gamma$, and system sizes $N$. We show a convergence at long times of the current when plotted against $\gamma^2 t/N$. This is a consequence of the perturbative character of the local coupling for larger systems; in fact the energy of the baths is extensive while the two baths are only coupled at a single site. 

When analyzing the current evolution over these time scales we can clearly recognize three regimes: a first transient regime, a regime in which one can observe (quasi-)steady currents for weak coupling (which we have shown to be typical), and a thermalization regime in which the current goes to zero. 
The emergence of the intermediate regime with a (quasi-)steady current is analogous to the phenomenon of prethermalization in isolated systems in which a thermalizing system with a weakly broken conserved law would relax slowly due to the underlying presence of conservation laws (the prethermalized regime), and then slowly relax toward the final states. In our case the conserved quantity is the energy of each bath, which is broken by the coupling between them. This dynamical correspondence with prethermalization is strengthened by the $\gamma^{2}$ relaxation rate \cite{MallayyaDeRoeck2019a, MallayyaRigol2021}.

\para{Conclusions} 
We studied the emergence of long-lasting currents between differently prepared nonintegrable systems in analogy to the emergence of prethermalization in isolated systems. Each bath has its own thermalizing dynamics which conserves its energy. This conservation law is weakly broken due to the coupling between the baths, and this results in a typical fast relaxation towards a nonequilibrium scenario with, in the weak-coupling regime, a (quasi-)steady current. This is then followed by a slow relaxation towards thermalization. We have shown that such dynamics is typical in the sense that it is quantitatively the same for any typical state chosen from two different energy shells in the respective baths, and specifically we have shown that this current is consistent with the prediction from ETH, and with preparing the baths in microcanonical states. The convergence towards ETH also results in a smooth dependence of current with energy difference. To obtain these results, we have used both perturbative and exact numerical methods. For the latter we used a highly parallelized algorithm which allowed us to simulate spin chains of the size of $28$ spins ($14$ spins per bath) and more. We have thus studied a model which makes it possible for the observation of nonequilibrium steady current as well as the long time thermalization. This connects the advances in equilibrium statistical mechanics with those in nonequilibrium open quantum systems. 

We would like to stress here that the study of transport and thermalization properties of systems by preparing two subsystems in different states and coupling them has already been undertaken in the past years \cite{PonomarevHanggi2011, BiellaFazio2016, MascarenhasSavona2017, BiellaMazza2019, LjubotinaProsen2017, ZnidaricLjubotina2018, LjubotinaProsen2019, BalachandranPoletti2018b}.     
With this work we lay a stronger basis on the application of these approaches, highlighting that the evolution of a relatively small number of different pure-state initial conditions can lead to accurate estimates of the currents expected from the microcanonical initial conditions. In the future we will consider adding an intermediate system between the baths so as to study conditions for the emergence of a (quasi-)steady current within this intermediate system. Furthermore, it would be important to consider whether the conditions for the emergence of the (quasi-)steady current can be loosened, for example considering also integrable baths \cite{BushongDiVentra2005, KarraschMoore2013, BiellaFazio2016, BrenesRigol2020, LeBlondRigol2020, LydzbaVidmar2021}.

\para{Acknowledgments} 
The authors are grateful to Ruofan Chen, Jochen Gemmer, Takaaki Monnai, Juzar Thingna, and Hangbo Zhou for fruitful discussions. D. P. acknowledges support from the Ministry of Education of Singapore AcRF MOE Tier-II (Project No. MOE2018-T2-2-142). C. G. acknowledges support from National Natural Science Foundation of China under Grants No. 11805279, No. 61833010, No. 12074117 and No. 12061131011. The computational work for this letter were partially performed on the National Supercomputing Centre, Singapore \cite{NSCC}. 


%


\begin{thebibliography}{67}%
\makeatletter
\providecommand \@ifxundefined [1]{%
 \@ifx{#1\undefined}
}%
\providecommand \@ifnum [1]{%
 \ifnum #1\expandafter \@firstoftwo
 \else \expandafter \@secondoftwo
 \fi
}%
\providecommand \@ifx [1]{%
 \ifx #1\expandafter \@firstoftwo
 \else \expandafter \@secondoftwo
 \fi
}%
\providecommand \natexlab [1]{#1}%
\providecommand \enquote  [1]{``#1''}%
\providecommand \bibnamefont  [1]{#1}%
\providecommand \bibfnamefont [1]{#1}%
\providecommand \citenamefont [1]{#1}%
\providecommand \href@noop [0]{\@secondoftwo}%
\providecommand \href [0]{\begingroup \@sanitize@url \@href}%
\providecommand \@href[1]{\@@startlink{#1}\@@href}%
\providecommand \@@href[1]{\endgroup#1\@@endlink}%
\providecommand \@sanitize@url [0]{\catcode `\\12\catcode `\$12\catcode
  `\&12\catcode `\#12\catcode `\^12\catcode `\_12\catcode `\%12\relax}%
\providecommand \@@startlink[1]{}%
\providecommand \@@endlink[0]{}%
\providecommand \url  [0]{\begingroup\@sanitize@url \@url }%
\providecommand \@url [1]{\endgroup\@href {#1}{\urlprefix }}%
\providecommand \urlprefix  [0]{URL }%
\providecommand \Eprint [0]{\href }%
\providecommand \doibase [0]{https://doi.org/}%
\providecommand \selectlanguage [0]{\@gobble}%
\providecommand \bibinfo  [0]{\@secondoftwo}%
\providecommand \bibfield  [0]{\@secondoftwo}%
\providecommand \translation [1]{[#1]}%
\providecommand \BibitemOpen [0]{}%
\providecommand \bibitemStop [0]{}%
\providecommand \bibitemNoStop [0]{.\EOS\space}%
\providecommand \EOS [0]{\spacefactor3000\relax}%
\providecommand \BibitemShut  [1]{\csname bibitem#1\endcsname}%
\let\auto@bib@innerbib\@empty
\bibitem [{\citenamefont {Gogolin}\ and\ \citenamefont
  {Eisert}(2016)}]{GogolinEisert2016}%
  \BibitemOpen
  \bibfield  {author} {\bibinfo {author} {\bibfnamefont {C.}~\bibnamefont
  {Gogolin}}\ and\ \bibinfo {author} {\bibfnamefont {J.}~\bibnamefont
  {Eisert}},\ }\bibfield  {title} {\bibinfo {title} {Equilibration,
  thermalisation, and the emergence of statistical mechanics in closed quantum
  systems},\ }\href {https://doi.org/10.1088/0034-4885/79/5/056001} {\bibfield
  {journal} {\bibinfo  {journal} {Rep. Prog. Phys.}\ }\textbf {\bibinfo
  {volume} {79}},\ \bibinfo {pages} {056001} (\bibinfo {year}
  {2016})}\BibitemShut {NoStop}%
\bibitem [{\citenamefont {Lloyd}(1988)}]{Lloyd1988}%
  \BibitemOpen
  \bibfield  {author} {\bibinfo {author} {\bibfnamefont {S.}~\bibnamefont
  {Lloyd}},\ }\emph {\bibinfo {title} {Black {{Holes}}, {{Demons}} and the
  {{Loss}} of {{Coherence}}: How Complex Systems Get Information, and What They
  Do with It.}},\ \href@noop {} {Ph.D. thesis},\ \bibinfo  {school} {The
  Rockefeller University} (\bibinfo {year} {1988})\BibitemShut {NoStop}%
\bibitem [{\citenamefont {Deutsch}(1991)}]{Deutsch1991}%
  \BibitemOpen
  \bibfield  {author} {\bibinfo {author} {\bibfnamefont {J.~M.}\ \bibnamefont
  {Deutsch}},\ }\bibfield  {title} {\bibinfo {title} {Quantum statistical
  mechanics in a closed system},\ }\href@noop {} {\bibfield  {journal}
  {\bibinfo  {journal} {Phys. Rev. A}\ }\textbf {\bibinfo {volume} {43}},\
  \bibinfo {pages} {2046} (\bibinfo {year} {1991})}\BibitemShut {NoStop}%
\bibitem [{\citenamefont {Srednicki}(1994)}]{Srednicki1994}%
  \BibitemOpen
  \bibfield  {author} {\bibinfo {author} {\bibfnamefont {M.}~\bibnamefont
  {Srednicki}},\ }\bibfield  {title} {\bibinfo {title} {Chaos and quantum
  thermalization},\ }\href {https://doi.org/10.1103/PhysRevE.50.888} {\bibfield
   {journal} {\bibinfo  {journal} {Phys. Rev. E}\ }\textbf {\bibinfo {volume}
  {50}},\ \bibinfo {pages} {888} (\bibinfo {year} {1994})}\BibitemShut
  {NoStop}%
\bibitem [{\citenamefont {Rigol}\ \emph {et~al.}(2008)\citenamefont {Rigol},
  \citenamefont {Dunjko},\ and\ \citenamefont {Olshanii}}]{RigolOlshanii2008}%
  \BibitemOpen
  \bibfield  {author} {\bibinfo {author} {\bibfnamefont {M.}~\bibnamefont
  {Rigol}}, \bibinfo {author} {\bibfnamefont {V.}~\bibnamefont {Dunjko}},\ and\
  \bibinfo {author} {\bibfnamefont {M.}~\bibnamefont {Olshanii}},\ }\bibfield
  {title} {\bibinfo {title} {Thermalization and its mechanism for generic
  isolated quantum systems},\ }\href {https://doi.org/10.1038/nature06838}
  {\bibfield  {journal} {\bibinfo  {journal} {Nature}\ }\textbf {\bibinfo
  {volume} {452}},\ \bibinfo {pages} {854} (\bibinfo {year}
  {2008})}\BibitemShut {NoStop}%
\bibitem [{\citenamefont {Borgonovi}\ \emph {et~al.}(2016)\citenamefont
  {Borgonovi}, \citenamefont {Izrailev}, \citenamefont {Santos},\ and\
  \citenamefont {Zelevinsky}}]{BorgonoviZelevinsky2016}%
  \BibitemOpen
  \bibfield  {author} {\bibinfo {author} {\bibfnamefont {F.}~\bibnamefont
  {Borgonovi}}, \bibinfo {author} {\bibfnamefont {F.}~\bibnamefont {Izrailev}},
  \bibinfo {author} {\bibfnamefont {L.}~\bibnamefont {Santos}},\ and\ \bibinfo
  {author} {\bibfnamefont {V.}~\bibnamefont {Zelevinsky}},\ }\bibfield  {title}
  {\bibinfo {title} {Quantum chaos and thermalization in isolated systems of
  interacting particles},\ }\href
  {https://doi.org/10.1016/j.physrep.2016.02.005} {\bibfield  {journal}
  {\bibinfo  {journal} {Phys. Rep.}\ }\textbf {\bibinfo {volume} {626}},\
  \bibinfo {pages} {1} (\bibinfo {year} {2016})}\BibitemShut {NoStop}%
\bibitem [{\citenamefont {D'Alessio}\ \emph {et~al.}(2016)\citenamefont
  {D'Alessio}, \citenamefont {Kafri}, \citenamefont {Polkovnikov},\ and\
  \citenamefont {Rigol}}]{DAlessioRigol2016}%
  \BibitemOpen
  \bibfield  {author} {\bibinfo {author} {\bibfnamefont {L.}~\bibnamefont
  {D'Alessio}}, \bibinfo {author} {\bibfnamefont {Y.}~\bibnamefont {Kafri}},
  \bibinfo {author} {\bibfnamefont {A.}~\bibnamefont {Polkovnikov}},\ and\
  \bibinfo {author} {\bibfnamefont {M.}~\bibnamefont {Rigol}},\ }\bibfield
  {title} {\bibinfo {title} {From quantum chaos and eigenstate thermalization
  to statistical mechanics and thermodynamics},\ }\href
  {https://doi.org/10.1080/00018732.2016.1198134} {\bibfield  {journal}
  {\bibinfo  {journal} {Adv. Phys.}\ }\textbf {\bibinfo {volume} {65}},\
  \bibinfo {pages} {239} (\bibinfo {year} {2016})}\BibitemShut {NoStop}%
\bibitem [{\citenamefont {Mori}\ \emph {et~al.}(2018)\citenamefont {Mori},
  \citenamefont {Ikeda}, \citenamefont {Kaminishi},\ and\ \citenamefont
  {Ueda}}]{MoriUeda2018}%
  \BibitemOpen
  \bibfield  {author} {\bibinfo {author} {\bibfnamefont {T.}~\bibnamefont
  {Mori}}, \bibinfo {author} {\bibfnamefont {T.~N.}\ \bibnamefont {Ikeda}},
  \bibinfo {author} {\bibfnamefont {E.}~\bibnamefont {Kaminishi}},\ and\
  \bibinfo {author} {\bibfnamefont {M.}~\bibnamefont {Ueda}},\ }\bibfield
  {title} {\bibinfo {title} {Thermalization and prethermalization in isolated
  quantum systems: A theoretical overview},\ }\href
  {https://doi.org/10.1088/1361-6455/aabcdf} {\bibfield  {journal} {\bibinfo
  {journal} {J. Phys. B: At. Mol. Opt. Phys.}\ }\textbf {\bibinfo {volume}
  {51}},\ \bibinfo {pages} {112001} (\bibinfo {year} {2018})}\BibitemShut
  {NoStop}%
\bibitem [{\citenamefont {Deutsch}(2018)}]{Deutsch2018}%
  \BibitemOpen
  \bibfield  {author} {\bibinfo {author} {\bibfnamefont {J.~M.}\ \bibnamefont
  {Deutsch}},\ }\bibfield  {title} {\bibinfo {title} {Eigenstate thermalization
  hypothesis},\ }\href {https://doi.org/10.1088/1361-6633/aac9f1} {\bibfield
  {journal} {\bibinfo  {journal} {Rep. Prog. Phys.}\ }\textbf {\bibinfo
  {volume} {81}},\ \bibinfo {pages} {082001} (\bibinfo {year}
  {2018})}\BibitemShut {NoStop}%
\bibitem [{\citenamefont {Ueda}(2020)}]{Ueda2020}%
  \BibitemOpen
  \bibfield  {author} {\bibinfo {author} {\bibfnamefont {M.}~\bibnamefont
  {Ueda}},\ }\bibfield  {title} {\bibinfo {title} {Quantum equilibration,
  thermalization and prethermalization in ultracold atoms},\ }\href
  {https://doi.org/10.1038/s42254-020-0237-x} {\bibfield  {journal} {\bibinfo
  {journal} {Nat. Rev. Phys.}\ }\textbf {\bibinfo {volume} {2}},\ \bibinfo
  {pages} {669} (\bibinfo {year} {2020})}\BibitemShut {NoStop}%
\bibitem [{\citenamefont {Reimann}(2007)}]{Reimann2007}%
  \BibitemOpen
  \bibfield  {author} {\bibinfo {author} {\bibfnamefont {P.}~\bibnamefont
  {Reimann}},\ }\bibfield  {title} {\bibinfo {title} {Typicality for
  {{Generalized Microcanonical Ensembles}}},\ }\href
  {https://doi.org/10.1103/PhysRevLett.99.160404} {\bibfield  {journal}
  {\bibinfo  {journal} {Phys. Rev. Lett.}\ }\textbf {\bibinfo {volume} {99}},\
  \bibinfo {pages} {160404} (\bibinfo {year} {2007})}\BibitemShut {NoStop}%
\bibitem [{\citenamefont {Bartsch}\ and\ \citenamefont
  {Gemmer}(2009)}]{BartschGemmer2009}%
  \BibitemOpen
  \bibfield  {author} {\bibinfo {author} {\bibfnamefont {C.}~\bibnamefont
  {Bartsch}}\ and\ \bibinfo {author} {\bibfnamefont {J.}~\bibnamefont
  {Gemmer}},\ }\bibfield  {title} {\bibinfo {title} {Dynamical {{Typicality}}
  of {{Quantum Expectation Values}}},\ }\href
  {https://doi.org/10.1103/PhysRevLett.102.110403} {\bibfield  {journal}
  {\bibinfo  {journal} {Phys. Rev. Lett.}\ }\textbf {\bibinfo {volume} {102}},\
  \bibinfo {pages} {110403} (\bibinfo {year} {2009})}\BibitemShut {NoStop}%
\bibitem [{\citenamefont {Gemmer}\ \emph {et~al.}(2010)\citenamefont {Gemmer},
  \citenamefont {Michel},\ and\ \citenamefont {Mahler}}]{GemmerMahler2010}%
  \BibitemOpen
  \bibfield  {author} {\bibinfo {author} {\bibfnamefont {J.}~\bibnamefont
  {Gemmer}}, \bibinfo {author} {\bibfnamefont {M.}~\bibnamefont {Michel}},\
  and\ \bibinfo {author} {\bibfnamefont {G.}~\bibnamefont {Mahler}},\ }\href
  {https://doi.org/10.1007/978-3-540-70510-9} {\emph {\bibinfo {title} {Quantum
  {{Thermodynamics}}}}},\ Vol.\ \bibinfo {volume} {784}\ (\bibinfo  {publisher}
  {{Springer Berlin Heidelberg}},\ \bibinfo {year} {2010})\BibitemShut
  {NoStop}%
\bibitem [{\citenamefont {Goldstein}\ \emph {et~al.}(2006)\citenamefont
  {Goldstein}, \citenamefont {Lebowitz}, \citenamefont {Tumulka},\ and\
  \citenamefont {Zangh{\`i}}}]{GoldsteinZanghi2006}%
  \BibitemOpen
  \bibfield  {author} {\bibinfo {author} {\bibfnamefont {S.}~\bibnamefont
  {Goldstein}}, \bibinfo {author} {\bibfnamefont {J.~L.}\ \bibnamefont
  {Lebowitz}}, \bibinfo {author} {\bibfnamefont {R.}~\bibnamefont {Tumulka}},\
  and\ \bibinfo {author} {\bibfnamefont {N.}~\bibnamefont {Zangh{\`i}}},\
  }\bibfield  {title} {\bibinfo {title} {Canonical {{Typicality}}},\ }\href
  {https://doi.org/10.1103/PhysRevLett.96.050403} {\bibfield  {journal}
  {\bibinfo  {journal} {Phys. Rev. Lett.}\ }\textbf {\bibinfo {volume} {96}},\
  \bibinfo {pages} {050403} (\bibinfo {year} {2006})}\BibitemShut {NoStop}%
\bibitem [{\citenamefont {Popescu}\ \emph {et~al.}(2006)\citenamefont
  {Popescu}, \citenamefont {Short},\ and\ \citenamefont
  {Winter}}]{PopescuWinter2006}%
  \BibitemOpen
  \bibfield  {author} {\bibinfo {author} {\bibfnamefont {S.}~\bibnamefont
  {Popescu}}, \bibinfo {author} {\bibfnamefont {A.~J.}\ \bibnamefont {Short}},\
  and\ \bibinfo {author} {\bibfnamefont {A.}~\bibnamefont {Winter}},\
  }\bibfield  {title} {\bibinfo {title} {Entanglement and the foundations of
  statistical mechanics},\ }\href {https://doi.org/10.1038/nphys444} {\bibfield
   {journal} {\bibinfo  {journal} {Nat. Phys.}\ }\textbf {\bibinfo {volume}
  {2}},\ \bibinfo {pages} {754} (\bibinfo {year} {2006})}\BibitemShut {NoStop}%
\bibitem [{\citenamefont {Reimann}(2018{\natexlab{a}})}]{Reimann2018a}%
  \BibitemOpen
  \bibfield  {author} {\bibinfo {author} {\bibfnamefont {P.}~\bibnamefont
  {Reimann}},\ }\bibfield  {title} {\bibinfo {title} {Dynamical typicality of
  isolated many-body quantum systems},\ }\href
  {https://doi.org/10.1103/PhysRevE.97.062129} {\bibfield  {journal} {\bibinfo
  {journal} {Phys. Rev. E}\ }\textbf {\bibinfo {volume} {97}},\ \bibinfo
  {pages} {062129} (\bibinfo {year} {2018}{\natexlab{a}})}\BibitemShut
  {NoStop}%
\bibitem [{\citenamefont {Reimann}(2018{\natexlab{b}})}]{Reimann2018}%
  \BibitemOpen
  \bibfield  {author} {\bibinfo {author} {\bibfnamefont {P.}~\bibnamefont
  {Reimann}},\ }\bibfield  {title} {\bibinfo {title} {Dynamical {{Typicality
  Approach}} to {{Eigenstate Thermalization}}},\ }\href
  {https://doi.org/10.1103/PhysRevLett.120.230601} {\bibfield  {journal}
  {\bibinfo  {journal} {Phys. Rev. Lett.}\ }\textbf {\bibinfo {volume} {120}},\
  \bibinfo {pages} {230601} (\bibinfo {year} {2018}{\natexlab{b}})}\BibitemShut
  {NoStop}%
\bibitem [{\citenamefont {Dabelow}\ and\ \citenamefont
  {Reimann}(2020)}]{DabelowReimann2020}%
  \BibitemOpen
  \bibfield  {author} {\bibinfo {author} {\bibfnamefont {L.}~\bibnamefont
  {Dabelow}}\ and\ \bibinfo {author} {\bibfnamefont {P.}~\bibnamefont
  {Reimann}},\ }\bibfield  {title} {\bibinfo {title} {Relaxation {{Theory}} for
  {{Perturbed Many}}-{{Body Quantum Systems}} versus {{Numerics}} and
  {{Experiment}}},\ }\href {https://doi.org/10.1103/PhysRevLett.124.120602}
  {\bibfield  {journal} {\bibinfo  {journal} {Phys. Rev. Lett.}\ }\textbf
  {\bibinfo {volume} {124}},\ \bibinfo {pages} {120602} (\bibinfo {year}
  {2020})}\BibitemShut {NoStop}%
\bibitem [{\citenamefont {Dabelow}\ and\ \citenamefont
  {Reimann}(2021)}]{DabelowReimann2021}%
  \BibitemOpen
  \bibfield  {author} {\bibinfo {author} {\bibfnamefont {L.}~\bibnamefont
  {Dabelow}}\ and\ \bibinfo {author} {\bibfnamefont {P.}~\bibnamefont
  {Reimann}},\ }\bibfield  {title} {\bibinfo {title} {Typical relaxation of
  perturbed quantum many-body systems},\ }\href
  {https://doi.org/10.1088/1742-5468/abd026} {\bibfield  {journal} {\bibinfo
  {journal} {J. Stat. Mech.}\ }\textbf {\bibinfo {volume} {2021}},\ \bibinfo
  {pages} {013106} (\bibinfo {year} {2021})}\BibitemShut {NoStop}%
\bibitem [{\citenamefont {Biroli}\ \emph {et~al.}(2010)\citenamefont {Biroli},
  \citenamefont {Kollath},\ and\ \citenamefont
  {L{\"a}uchli}}]{BiroliLauchli2010}%
  \BibitemOpen
  \bibfield  {author} {\bibinfo {author} {\bibfnamefont {G.}~\bibnamefont
  {Biroli}}, \bibinfo {author} {\bibfnamefont {C.}~\bibnamefont {Kollath}},\
  and\ \bibinfo {author} {\bibfnamefont {A.~M.}\ \bibnamefont {L{\"a}uchli}},\
  }\bibfield  {title} {\bibinfo {title} {Effect of {{Rare Fluctuations}} on the
  {{Thermalization}} of {{Isolated Quantum Systems}}},\ }\href
  {https://doi.org/10.1103/PhysRevLett.105.250401} {\bibfield  {journal}
  {\bibinfo  {journal} {Phys. Rev. Lett.}\ }\textbf {\bibinfo {volume} {105}},\
  \bibinfo {pages} {250401} (\bibinfo {year} {2010})}\BibitemShut {NoStop}%
\bibitem [{\citenamefont {Ikeda}\ \emph {et~al.}(2013)\citenamefont {Ikeda},
  \citenamefont {Watanabe},\ and\ \citenamefont {Ueda}}]{IkedaUeda2013}%
  \BibitemOpen
  \bibfield  {author} {\bibinfo {author} {\bibfnamefont {T.~N.}\ \bibnamefont
  {Ikeda}}, \bibinfo {author} {\bibfnamefont {Y.}~\bibnamefont {Watanabe}},\
  and\ \bibinfo {author} {\bibfnamefont {M.}~\bibnamefont {Ueda}},\ }\bibfield
  {title} {\bibinfo {title} {Finite-size scaling analysis of the eigenstate
  thermalization hypothesis in a one-dimensional interacting {{Bose}} gas},\
  }\href {https://doi.org/10.1103/PhysRevE.87.012125} {\bibfield  {journal}
  {\bibinfo  {journal} {Phys. Rev. E}\ }\textbf {\bibinfo {volume} {87}},\
  \bibinfo {pages} {012125} (\bibinfo {year} {2013})}\BibitemShut {NoStop}%
\bibitem [{\citenamefont {Beugeling}\ \emph {et~al.}(2014)\citenamefont
  {Beugeling}, \citenamefont {Moessner},\ and\ \citenamefont
  {Haque}}]{BeugelingHaque2014}%
  \BibitemOpen
  \bibfield  {author} {\bibinfo {author} {\bibfnamefont {W.}~\bibnamefont
  {Beugeling}}, \bibinfo {author} {\bibfnamefont {R.}~\bibnamefont
  {Moessner}},\ and\ \bibinfo {author} {\bibfnamefont {M.}~\bibnamefont
  {Haque}},\ }\bibfield  {title} {\bibinfo {title} {Finite-size scaling of
  eigenstate thermalization},\ }\href
  {https://doi.org/10.1103/PhysRevE.89.042112} {\bibfield  {journal} {\bibinfo
  {journal} {Phys. Rev. E}\ }\textbf {\bibinfo {volume} {89}},\ \bibinfo
  {pages} {042112} (\bibinfo {year} {2014})}\BibitemShut {NoStop}%
\bibitem [{\citenamefont {Iyoda}\ \emph {et~al.}(2017)\citenamefont {Iyoda},
  \citenamefont {Kaneko},\ and\ \citenamefont {Sagawa}}]{IyodaSagawa2017}%
  \BibitemOpen
  \bibfield  {author} {\bibinfo {author} {\bibfnamefont {E.}~\bibnamefont
  {Iyoda}}, \bibinfo {author} {\bibfnamefont {K.}~\bibnamefont {Kaneko}},\ and\
  \bibinfo {author} {\bibfnamefont {T.}~\bibnamefont {Sagawa}},\ }\bibfield
  {title} {\bibinfo {title} {Fluctuation {{Theorem}} for {{Many}}-{{Body Pure
  Quantum States}}},\ }\href {https://doi.org/10.1103/PhysRevLett.119.100601}
  {\bibfield  {journal} {\bibinfo  {journal} {Phys. Rev. Lett.}\ }\textbf
  {\bibinfo {volume} {119}},\ \bibinfo {pages} {100601} (\bibinfo {year}
  {2017})}\BibitemShut {NoStop}%
\bibitem [{\citenamefont {Yoshizawa}\ \emph {et~al.}(2018)\citenamefont
  {Yoshizawa}, \citenamefont {Iyoda},\ and\ \citenamefont
  {Sagawa}}]{YoshizawaSagawa2018}%
  \BibitemOpen
  \bibfield  {author} {\bibinfo {author} {\bibfnamefont {T.}~\bibnamefont
  {Yoshizawa}}, \bibinfo {author} {\bibfnamefont {E.}~\bibnamefont {Iyoda}},\
  and\ \bibinfo {author} {\bibfnamefont {T.}~\bibnamefont {Sagawa}},\
  }\bibfield  {title} {\bibinfo {title} {Numerical {{Large Deviation Analysis}}
  of the {{Eigenstate Thermalization Hypothesis}}},\ }\href
  {https://doi.org/10.1103/PhysRevLett.120.200604} {\bibfield  {journal}
  {\bibinfo  {journal} {Phys. Rev. Lett.}\ }\textbf {\bibinfo {volume} {120}},\
  \bibinfo {pages} {200604} (\bibinfo {year} {2018})}\BibitemShut {NoStop}%
\bibitem [{\citenamefont {Berges}\ \emph {et~al.}(2004)\citenamefont {Berges},
  \citenamefont {Borsanyi},\ and\ \citenamefont
  {Wetterich}}]{BergesWetterich2004}%
  \BibitemOpen
  \bibfield  {author} {\bibinfo {author} {\bibfnamefont {J.}~\bibnamefont
  {Berges}}, \bibinfo {author} {\bibfnamefont {S.}~\bibnamefont {Borsanyi}},\
  and\ \bibinfo {author} {\bibfnamefont {C.}~\bibnamefont {Wetterich}},\
  }\bibfield  {title} {\bibinfo {title} {Prethermalization},\ }\href
  {https://doi.org/10.1103/PhysRevLett.93.142002} {\bibfield  {journal}
  {\bibinfo  {journal} {Phys. Rev. Lett.}\ }\textbf {\bibinfo {volume} {93}},\
  \bibinfo {pages} {142002} (\bibinfo {year} {2004})}\BibitemShut {NoStop}%
\bibitem [{\citenamefont {Moeckel}\ and\ \citenamefont
  {Kehrein}(2008)}]{MoeckelKehrein2008}%
  \BibitemOpen
  \bibfield  {author} {\bibinfo {author} {\bibfnamefont {M.}~\bibnamefont
  {Moeckel}}\ and\ \bibinfo {author} {\bibfnamefont {S.}~\bibnamefont
  {Kehrein}},\ }\bibfield  {title} {\bibinfo {title} {Interaction {{Quench}} in
  the {{Hubbard Model}}},\ }\href
  {https://doi.org/10.1103/PhysRevLett.100.175702} {\bibfield  {journal}
  {\bibinfo  {journal} {Phys. Rev. Lett.}\ }\textbf {\bibinfo {volume} {100}},\
  \bibinfo {pages} {175702} (\bibinfo {year} {2008})}\BibitemShut {NoStop}%
\bibitem [{\citenamefont {Eckstein}\ \emph {et~al.}(2010)\citenamefont
  {Eckstein}, \citenamefont {Kollar},\ and\ \citenamefont
  {Werner}}]{EcksteinWerner2010}%
  \BibitemOpen
  \bibfield  {author} {\bibinfo {author} {\bibfnamefont {M.}~\bibnamefont
  {Eckstein}}, \bibinfo {author} {\bibfnamefont {M.}~\bibnamefont {Kollar}},\
  and\ \bibinfo {author} {\bibfnamefont {P.}~\bibnamefont {Werner}},\
  }\bibfield  {title} {\bibinfo {title} {Interaction quench in the {{Hubbard}}
  model: Relaxation of the spectral function and the optical conductivity},\
  }\href {https://doi.org/10.1103/PhysRevB.81.115131} {\bibfield  {journal}
  {\bibinfo  {journal} {Phys. Rev. B}\ }\textbf {\bibinfo {volume} {81}},\
  \bibinfo {pages} {115131} (\bibinfo {year} {2010})}\BibitemShut {NoStop}%
\bibitem [{\citenamefont {Kollar}\ \emph {et~al.}(2011)\citenamefont {Kollar},
  \citenamefont {Wolf},\ and\ \citenamefont {Eckstein}}]{KollarEckstein2011}%
  \BibitemOpen
  \bibfield  {author} {\bibinfo {author} {\bibfnamefont {M.}~\bibnamefont
  {Kollar}}, \bibinfo {author} {\bibfnamefont {F.~A.}\ \bibnamefont {Wolf}},\
  and\ \bibinfo {author} {\bibfnamefont {M.}~\bibnamefont {Eckstein}},\
  }\bibfield  {title} {\bibinfo {title} {Generalized {{Gibbs}} ensemble
  prediction of prethermalization plateaus and their relation to nonthermal
  steady states in integrable systems},\ }\href
  {https://doi.org/10.1103/PhysRevB.84.054304} {\bibfield  {journal} {\bibinfo
  {journal} {Phys. Rev. B}\ }\textbf {\bibinfo {volume} {84}},\ \bibinfo
  {pages} {054304} (\bibinfo {year} {2011})}\BibitemShut {NoStop}%
\bibitem [{\citenamefont {Aoki}\ \emph {et~al.}(2014)\citenamefont {Aoki},
  \citenamefont {Tsuji}, \citenamefont {Eckstein}, \citenamefont {Kollar},
  \citenamefont {Oka},\ and\ \citenamefont {Werner}}]{AokiWerner2014}%
  \BibitemOpen
  \bibfield  {author} {\bibinfo {author} {\bibfnamefont {H.}~\bibnamefont
  {Aoki}}, \bibinfo {author} {\bibfnamefont {N.}~\bibnamefont {Tsuji}},
  \bibinfo {author} {\bibfnamefont {M.}~\bibnamefont {Eckstein}}, \bibinfo
  {author} {\bibfnamefont {M.}~\bibnamefont {Kollar}}, \bibinfo {author}
  {\bibfnamefont {T.}~\bibnamefont {Oka}},\ and\ \bibinfo {author}
  {\bibfnamefont {P.}~\bibnamefont {Werner}},\ }\bibfield  {title} {\bibinfo
  {title} {Nonequilibrium dynamical mean-field theory and its applications},\
  }\href {https://doi.org/10.1103/RevModPhys.86.779} {\bibfield  {journal}
  {\bibinfo  {journal} {Rev. Mod. Phys.}\ }\textbf {\bibinfo {volume} {86}},\
  \bibinfo {pages} {779} (\bibinfo {year} {2014})}\BibitemShut {NoStop}%
\bibitem [{\citenamefont {Mallayya}\ \emph {et~al.}(2019)\citenamefont
  {Mallayya}, \citenamefont {Rigol},\ and\ \citenamefont
  {De~Roeck}}]{MallayyaDeRoeck2019a}%
  \BibitemOpen
  \bibfield  {author} {\bibinfo {author} {\bibfnamefont {K.}~\bibnamefont
  {Mallayya}}, \bibinfo {author} {\bibfnamefont {M.}~\bibnamefont {Rigol}},\
  and\ \bibinfo {author} {\bibfnamefont {W.}~\bibnamefont {De~Roeck}},\
  }\bibfield  {title} {\bibinfo {title} {Prethermalization and
  {{Thermalization}} in {{Isolated Quantum Systems}}},\ }\href
  {https://doi.org/10.1103/PhysRevX.9.021027} {\bibfield  {journal} {\bibinfo
  {journal} {Phys. Rev. X}\ }\textbf {\bibinfo {volume} {9}},\ \bibinfo {pages}
  {021027} (\bibinfo {year} {2019})}\BibitemShut {NoStop}%
\bibitem [{\citenamefont {Mallayya}\ and\ \citenamefont
  {Rigol}(2021)}]{MallayyaRigol2021}%
  \BibitemOpen
  \bibfield  {author} {\bibinfo {author} {\bibfnamefont {K.}~\bibnamefont
  {Mallayya}}\ and\ \bibinfo {author} {\bibfnamefont {M.}~\bibnamefont
  {Rigol}},\ }\bibfield  {title} {\bibinfo {title} {Prethermalization,
  thermalization, and {{Fermi}}'s golden rule in quantum many-body systems},\
  }\href@noop {} {\bibfield  {journal} {\bibinfo  {journal} {arXiv:2109.01705}\
  } (\bibinfo {year} {2021})}\BibitemShut {NoStop}%
\bibitem [{\citenamefont {Gring}\ \emph {et~al.}(2012)\citenamefont {Gring},
  \citenamefont {Kuhnert}, \citenamefont {Langen}, \citenamefont {Kitagawa},
  \citenamefont {Rauer}, \citenamefont {Schreitl}, \citenamefont {Mazets},
  \citenamefont {Smith}, \citenamefont {Demler},\ and\ \citenamefont
  {Schmiedmayer}}]{GringSchmiedmayer2012}%
  \BibitemOpen
  \bibfield  {author} {\bibinfo {author} {\bibfnamefont {M.}~\bibnamefont
  {Gring}}, \bibinfo {author} {\bibfnamefont {M.}~\bibnamefont {Kuhnert}},
  \bibinfo {author} {\bibfnamefont {T.}~\bibnamefont {Langen}}, \bibinfo
  {author} {\bibfnamefont {T.}~\bibnamefont {Kitagawa}}, \bibinfo {author}
  {\bibfnamefont {B.}~\bibnamefont {Rauer}}, \bibinfo {author} {\bibfnamefont
  {M.}~\bibnamefont {Schreitl}}, \bibinfo {author} {\bibfnamefont
  {I.}~\bibnamefont {Mazets}}, \bibinfo {author} {\bibfnamefont {D.~A.}\
  \bibnamefont {Smith}}, \bibinfo {author} {\bibfnamefont {E.}~\bibnamefont
  {Demler}},\ and\ \bibinfo {author} {\bibfnamefont {J.}~\bibnamefont
  {Schmiedmayer}},\ }\bibfield  {title} {\bibinfo {title} {Relaxation and
  {{Prethermalization}} in an {{Isolated Quantum System}}},\ }\href
  {https://doi.org/10.1126/science.1224953} {\bibfield  {journal} {\bibinfo
  {journal} {Science}\ }\textbf {\bibinfo {volume} {337}},\ \bibinfo {pages}
  {1318} (\bibinfo {year} {2012})}\BibitemShut {NoStop}%
\bibitem [{\citenamefont {Reimann}\ and\ \citenamefont
  {Dabelow}(2019)}]{ReimannDabelow2019a}%
  \BibitemOpen
  \bibfield  {author} {\bibinfo {author} {\bibfnamefont {P.}~\bibnamefont
  {Reimann}}\ and\ \bibinfo {author} {\bibfnamefont {L.}~\bibnamefont
  {Dabelow}},\ }\bibfield  {title} {\bibinfo {title} {Typicality of
  {{Prethermalization}}},\ }\href
  {https://doi.org/10.1103/PhysRevLett.122.080603} {\bibfield  {journal}
  {\bibinfo  {journal} {Phys. Rev. Lett.}\ }\textbf {\bibinfo {volume} {122}},\
  \bibinfo {pages} {080603} (\bibinfo {year} {2019})}\BibitemShut {NoStop}%
\bibitem [{\citenamefont {Monnai}\ and\ \citenamefont
  {Yuasa}(2014)}]{MonnaiYuasa2014}%
  \BibitemOpen
  \bibfield  {author} {\bibinfo {author} {\bibfnamefont {T.}~\bibnamefont
  {Monnai}}\ and\ \bibinfo {author} {\bibfnamefont {K.}~\bibnamefont {Yuasa}},\
  }\bibfield  {title} {\bibinfo {title} {Typical pure nonequilibrium steady
  states},\ }\href {https://doi.org/10.1209/0295-5075/107/40006} {\bibfield
  {journal} {\bibinfo  {journal} {Europhys. Lett.}\ }\textbf {\bibinfo {volume}
  {107}},\ \bibinfo {pages} {40006} (\bibinfo {year} {2014})}\BibitemShut
  {NoStop}%
\bibitem [{\citenamefont {Monnai}\ and\ \citenamefont
  {Yuasa}(2016)}]{MonnaiYuasa2016}%
  \BibitemOpen
  \bibfield  {author} {\bibinfo {author} {\bibfnamefont {T.}~\bibnamefont
  {Monnai}}\ and\ \bibinfo {author} {\bibfnamefont {K.}~\bibnamefont {Yuasa}},\
  }\bibfield  {title} {\bibinfo {title} {Typical pure nonequilibrium steady
  states and irreversibility for quantum transport},\ }\href
  {https://doi.org/10.1103/PhysRevE.94.012146} {\bibfield  {journal} {\bibinfo
  {journal} {Phys. Rev. E}\ }\textbf {\bibinfo {volume} {94}},\ \bibinfo
  {pages} {012146} (\bibinfo {year} {2016})}\BibitemShut {NoStop}%
\bibitem [{\citenamefont {Evans}\ \emph {et~al.}(2016)\citenamefont {Evans},
  \citenamefont {Williams}, \citenamefont {Searles},\ and\ \citenamefont
  {Rondoni}}]{EvansRondoni2016}%
  \BibitemOpen
  \bibfield  {author} {\bibinfo {author} {\bibfnamefont {D.~J.}\ \bibnamefont
  {Evans}}, \bibinfo {author} {\bibfnamefont {S.~R.}\ \bibnamefont {Williams}},
  \bibinfo {author} {\bibfnamefont {D.~J.}\ \bibnamefont {Searles}},\ and\
  \bibinfo {author} {\bibfnamefont {L.}~\bibnamefont {Rondoni}},\ }\bibfield
  {title} {\bibinfo {title} {On {{Typicality}} in {{Nonequilibrium Steady
  States}}},\ }\href {https://doi.org/10.1007/s10955-016-1563-3} {\bibfield
  {journal} {\bibinfo  {journal} {J. Stat. Phys.}\ }\textbf {\bibinfo {volume}
  {164}},\ \bibinfo {pages} {842} (\bibinfo {year} {2016})}\BibitemShut
  {NoStop}%
\bibitem [{\citenamefont {Moudgalya}\ \emph {et~al.}(2019)\citenamefont
  {Moudgalya}, \citenamefont {Devakul}, \citenamefont {Arovas},\ and\
  \citenamefont {Sondhi}}]{MoudgalyaSondhi2019a}%
  \BibitemOpen
  \bibfield  {author} {\bibinfo {author} {\bibfnamefont {S.}~\bibnamefont
  {Moudgalya}}, \bibinfo {author} {\bibfnamefont {T.}~\bibnamefont {Devakul}},
  \bibinfo {author} {\bibfnamefont {D.~P.}\ \bibnamefont {Arovas}},\ and\
  \bibinfo {author} {\bibfnamefont {S.~L.}\ \bibnamefont {Sondhi}},\ }\bibfield
   {title} {\bibinfo {title} {Extension of the eigenstate thermalization
  hypothesis to nonequilibrium steady states},\ }\href
  {https://doi.org/10.1103/PhysRevB.100.045112} {\bibfield  {journal} {\bibinfo
   {journal} {Phys. Rev. B}\ }\textbf {\bibinfo {volume} {100}},\ \bibinfo
  {pages} {045112} (\bibinfo {year} {2019})}\BibitemShut {NoStop}%
\bibitem [{\citenamefont {Shirai}\ and\ \citenamefont
  {Mori}(2020)}]{ShiraiMori2020}%
  \BibitemOpen
  \bibfield  {author} {\bibinfo {author} {\bibfnamefont {T.}~\bibnamefont
  {Shirai}}\ and\ \bibinfo {author} {\bibfnamefont {T.}~\bibnamefont {Mori}},\
  }\bibfield  {title} {\bibinfo {title} {Thermalization in open many-body
  systems based on eigenstate thermalization hypothesis},\ }\href
  {https://doi.org/10.1103/PhysRevE.101.042116} {\bibfield  {journal} {\bibinfo
   {journal} {Phys. Rev. E}\ }\textbf {\bibinfo {volume} {101}},\ \bibinfo
  {pages} {042116} (\bibinfo {year} {2020})}\BibitemShut {NoStop}%
\bibitem [{\citenamefont {Atas}\ \emph {et~al.}(2013)\citenamefont {Atas},
  \citenamefont {Bogomolny}, \citenamefont {Giraud},\ and\ \citenamefont
  {Roux}}]{AtasRoux2013}%
  \BibitemOpen
  \bibfield  {author} {\bibinfo {author} {\bibfnamefont {Y.~Y.}\ \bibnamefont
  {Atas}}, \bibinfo {author} {\bibfnamefont {E.}~\bibnamefont {Bogomolny}},
  \bibinfo {author} {\bibfnamefont {O.}~\bibnamefont {Giraud}},\ and\ \bibinfo
  {author} {\bibfnamefont {G.}~\bibnamefont {Roux}},\ }\bibfield  {title}
  {\bibinfo {title} {Distribution of the {{Ratio}} of {{Consecutive Level
  Spacings}} in {{Random Matrix Ensembles}}},\ }\href
  {https://doi.org/10.1103/PhysRevLett.110.084101} {\bibfield  {journal}
  {\bibinfo  {journal} {Phys. Rev. Lett.}\ }\textbf {\bibinfo {volume} {110}},\
  \bibinfo {pages} {084101} (\bibinfo {year} {2013})}\BibitemShut {NoStop}%
\bibitem [{\citenamefont {Huang}\ \emph {et~al.}(2019)\citenamefont {Huang},
  \citenamefont {Brand{\~a}o},\ and\ \citenamefont {Zhang}}]{HuangZhang2019}%
  \BibitemOpen
  \bibfield  {author} {\bibinfo {author} {\bibfnamefont {Y.}~\bibnamefont
  {Huang}}, \bibinfo {author} {\bibfnamefont {F.~G. S.~L.}\ \bibnamefont
  {Brand{\~a}o}},\ and\ \bibinfo {author} {\bibfnamefont {Y.-L.}\ \bibnamefont
  {Zhang}},\ }\bibfield  {title} {\bibinfo {title} {Finite-{{Size Scaling}} of
  {{Out}}-of-{{Time}}-{{Ordered Correlators}} at {{Late Times}}},\ }\href
  {https://doi.org/10.1103/PhysRevLett.123.010601} {\bibfield  {journal}
  {\bibinfo  {journal} {Phys. Rev. Lett.}\ }\textbf {\bibinfo {volume} {123}},\
  \bibinfo {pages} {010601} (\bibinfo {year} {2019})}\BibitemShut {NoStop}%
\bibitem [{Note1()}]{Note1}%
  \BibitemOpen
  \bibinfo {note} {Note that when discussing the expectation value of the
  current we drop the superscript ${\protect \rm L}$ because, in the
  (quasi-)steady regime we are interested in, the current from the left bath
  equals that to the right bath. More details in \cite {Supp}}\BibitemShut
  {NoStop}%
\bibitem [{\citenamefont {Rigol}\ \emph {et~al.}(2006)\citenamefont {Rigol},
  \citenamefont {Bryant},\ and\ \citenamefont {Singh}}]{RigolSingh2006}%
  \BibitemOpen
  \bibfield  {author} {\bibinfo {author} {\bibfnamefont {M.}~\bibnamefont
  {Rigol}}, \bibinfo {author} {\bibfnamefont {T.}~\bibnamefont {Bryant}},\ and\
  \bibinfo {author} {\bibfnamefont {R.~R.~P.}\ \bibnamefont {Singh}},\
  }\bibfield  {title} {\bibinfo {title} {Numerical {{Linked}}-{{Cluster
  Approach}} to {{Quantum Lattice Models}}},\ }\href
  {https://doi.org/10.1103/PhysRevLett.97.187202} {\bibfield  {journal}
  {\bibinfo  {journal} {Phys. Rev. Lett.}\ }\textbf {\bibinfo {volume} {97}},\
  \bibinfo {pages} {187202} (\bibinfo {year} {2006})}\BibitemShut {NoStop}%
\bibitem [{\citenamefont {Steinigeweg}\ \emph {et~al.}(2014)\citenamefont
  {Steinigeweg}, \citenamefont {Khodja}, \citenamefont {Niemeyer},
  \citenamefont {Gogolin},\ and\ \citenamefont
  {Gemmer}}]{SteinigewegGemmer2014}%
  \BibitemOpen
  \bibfield  {author} {\bibinfo {author} {\bibfnamefont {R.}~\bibnamefont
  {Steinigeweg}}, \bibinfo {author} {\bibfnamefont {A.}~\bibnamefont {Khodja}},
  \bibinfo {author} {\bibfnamefont {H.}~\bibnamefont {Niemeyer}}, \bibinfo
  {author} {\bibfnamefont {C.}~\bibnamefont {Gogolin}},\ and\ \bibinfo {author}
  {\bibfnamefont {J.}~\bibnamefont {Gemmer}},\ }\bibfield  {title} {\bibinfo
  {title} {Pushing the {{Limits}} of the {{Eigenstate Thermalization
  Hypothesis}} towards {{Mesoscopic Quantum Systems}}},\ }\href
  {https://doi.org/10.1103/PhysRevLett.112.130403} {\bibfield  {journal}
  {\bibinfo  {journal} {Phys. Rev. Lett.}\ }\textbf {\bibinfo {volume} {112}},\
  \bibinfo {pages} {130403} (\bibinfo {year} {2014})}\BibitemShut {NoStop}%
\bibitem [{Sup()}]{Supp}%
  \BibitemOpen
  \href@noop {} {\bibinfo {title} {See {{Supplementary}} material}}\BibitemShut
  {NoStop}%
\bibitem [{\citenamefont {Esposito}\ and\ \citenamefont
  {Gaspard}(2003)}]{EspositoGaspard2003b}%
  \BibitemOpen
  \bibfield  {author} {\bibinfo {author} {\bibfnamefont {M.}~\bibnamefont
  {Esposito}}\ and\ \bibinfo {author} {\bibfnamefont {P.}~\bibnamefont
  {Gaspard}},\ }\bibfield  {title} {\bibinfo {title} {Quantum master equation
  for a system influencing its environment},\ }\href
  {https://doi.org/10.1103/PhysRevE.68.066112} {\bibfield  {journal} {\bibinfo
  {journal} {Phys. Rev. E}\ }\textbf {\bibinfo {volume} {68}},\ \bibinfo
  {pages} {066112} (\bibinfo {year} {2003})}\BibitemShut {NoStop}%
\bibitem [{\citenamefont {Esposito}\ and\ \citenamefont
  {Gaspard}(2007)}]{EspositoGaspard2007}%
  \BibitemOpen
  \bibfield  {author} {\bibinfo {author} {\bibfnamefont {M.}~\bibnamefont
  {Esposito}}\ and\ \bibinfo {author} {\bibfnamefont {P.}~\bibnamefont
  {Gaspard}},\ }\bibfield  {title} {\bibinfo {title} {Quantum master equation
  for the microcanonical ensemble},\ }\href
  {https://doi.org/10.1103/PhysRevE.76.041134} {\bibfield  {journal} {\bibinfo
  {journal} {Phys. Rev. E}\ }\textbf {\bibinfo {volume} {76}},\ \bibinfo
  {pages} {041134} (\bibinfo {year} {2007})}\BibitemShut {NoStop}%
\bibitem [{\citenamefont {Breuer}\ and\ \citenamefont
  {Petruccione}(2007)}]{BreuerPetruccione2007}%
  \BibitemOpen
  \bibfield  {author} {\bibinfo {author} {\bibfnamefont {H.-P.}\ \bibnamefont
  {Breuer}}\ and\ \bibinfo {author} {\bibfnamefont {F.}~\bibnamefont
  {Petruccione}},\ }\href@noop {} {\emph {\bibinfo {title} {The {{Theory}} of
  {{Open Quantum Systems}}}}}\ (\bibinfo  {publisher} {{Oxford University
  Press}},\ \bibinfo {year} {2007})\BibitemShut {NoStop}%
\bibitem [{\citenamefont {{de Vega}}\ and\ \citenamefont
  {Alonso}(2017)}]{DeVegaAlonso2017}%
  \BibitemOpen
  \bibfield  {author} {\bibinfo {author} {\bibfnamefont {I.}~\bibnamefont {{de
  Vega}}}\ and\ \bibinfo {author} {\bibfnamefont {D.}~\bibnamefont {Alonso}},\
  }\bibfield  {title} {\bibinfo {title} {Dynamics of non-{{Markovian}} open
  quantum systems},\ }\href {https://doi.org/10.1103/RevModPhys.89.015001}
  {\bibfield  {journal} {\bibinfo  {journal} {Rev. Mod. Phys.}\ }\textbf
  {\bibinfo {volume} {89}},\ \bibinfo {pages} {015001} (\bibinfo {year}
  {2017})}\BibitemShut {NoStop}%
\bibitem [{\citenamefont {Landi}\ \emph {et~al.}(2021)\citenamefont {Landi},
  \citenamefont {Poletti},\ and\ \citenamefont {Schaller}}]{LandiSchaller2021}%
  \BibitemOpen
  \bibfield  {author} {\bibinfo {author} {\bibfnamefont {G.~T.}\ \bibnamefont
  {Landi}}, \bibinfo {author} {\bibfnamefont {D.}~\bibnamefont {Poletti}},\
  and\ \bibinfo {author} {\bibfnamefont {G.}~\bibnamefont {Schaller}},\
  }\bibfield  {title} {\bibinfo {title} {Non-equilibrium boundary driven
  quantum systems: Models, methods and properties},\ }\href@noop {} {\bibfield
  {journal} {\bibinfo  {journal} {arXiv:2104.14350}\ } (\bibinfo {year}
  {2021})}\BibitemShut {NoStop}%
\bibitem [{\citenamefont {Zhou}\ \emph {et~al.}(2020)\citenamefont {Zhou},
  \citenamefont {Zhang}, \citenamefont {Wang},\ and\ \citenamefont
  {Zhang}}]{ZhouZhang2020}%
  \BibitemOpen
  \bibfield  {author} {\bibinfo {author} {\bibfnamefont {H.}~\bibnamefont
  {Zhou}}, \bibinfo {author} {\bibfnamefont {G.}~\bibnamefont {Zhang}},
  \bibinfo {author} {\bibfnamefont {J.-S.}\ \bibnamefont {Wang}},\ and\
  \bibinfo {author} {\bibfnamefont {Y.-W.}\ \bibnamefont {Zhang}},\ }\bibfield
  {title} {\bibinfo {title} {Three-terminal interface as a thermoelectric
  generator beyond the {{Seebeck}} effect},\ }\href
  {https://doi.org/10.1103/PhysRevB.101.235305} {\bibfield  {journal} {\bibinfo
   {journal} {Phys. Rev. B}\ }\textbf {\bibinfo {volume} {101}},\ \bibinfo
  {pages} {235305} (\bibinfo {year} {2020})}\BibitemShut {NoStop}%
\bibitem [{Note2()}]{Note2}%
  \BibitemOpen
  \bibinfo {note} {The evolution of the currents from the corresponding
  different typical initial conditions are shown in \cite {Supp} for time up to
  $t=50$}\BibitemShut {NoStop}%
\bibitem [{\citenamefont {Srednicki}(1996)}]{Srednicki1996}%
  \BibitemOpen
  \bibfield  {author} {\bibinfo {author} {\bibfnamefont {M.}~\bibnamefont
  {Srednicki}},\ }\bibfield  {title} {\bibinfo {title} {Thermal fluctuations in
  quantized chaotic systems},\ }\href
  {https://doi.org/10.1088/0305-4470/29/4/003} {\bibfield  {journal} {\bibinfo
  {journal} {J. Phys. A: Math. Gen.}\ }\textbf {\bibinfo {volume} {29}},\
  \bibinfo {pages} {L75} (\bibinfo {year} {1996})}\BibitemShut {NoStop}%
\bibitem [{\citenamefont {Luitz}\ and\ \citenamefont
  {Bar~Lev}(2016)}]{LuitzBarLev2016}%
  \BibitemOpen
  \bibfield  {author} {\bibinfo {author} {\bibfnamefont {D.~J.}\ \bibnamefont
  {Luitz}}\ and\ \bibinfo {author} {\bibfnamefont {Y.}~\bibnamefont
  {Bar~Lev}},\ }\bibfield  {title} {\bibinfo {title} {Anomalous
  {{Thermalization}} in {{Ergodic Systems}}},\ }\href
  {https://doi.org/10.1103/PhysRevLett.117.170404} {\bibfield  {journal}
  {\bibinfo  {journal} {Phys. Rev. Lett.}\ }\textbf {\bibinfo {volume} {117}},\
  \bibinfo {pages} {170404} (\bibinfo {year} {2016})}\BibitemShut {NoStop}%
\bibitem [{\citenamefont {Ponomarev}\ \emph {et~al.}(2011)\citenamefont
  {Ponomarev}, \citenamefont {Denisov},\ and\ \citenamefont
  {H{\"a}nggi}}]{PonomarevHanggi2011}%
  \BibitemOpen
  \bibfield  {author} {\bibinfo {author} {\bibfnamefont {A.~V.}\ \bibnamefont
  {Ponomarev}}, \bibinfo {author} {\bibfnamefont {S.}~\bibnamefont {Denisov}},\
  and\ \bibinfo {author} {\bibfnamefont {P.}~\bibnamefont {H{\"a}nggi}},\
  }\bibfield  {title} {\bibinfo {title} {Thermal {{Equilibration}} between
  {{Two Quantum Systems}}},\ }\href
  {https://doi.org/10.1103/PhysRevLett.106.010405} {\bibfield  {journal}
  {\bibinfo  {journal} {Phys. Rev. Lett.}\ }\textbf {\bibinfo {volume} {106}},\
  \bibinfo {pages} {010405} (\bibinfo {year} {2011})}\BibitemShut {NoStop}%
\bibitem [{\citenamefont {Biella}\ \emph {et~al.}(2016)\citenamefont {Biella},
  \citenamefont {De~Luca}, \citenamefont {Viti}, \citenamefont {Rossini},
  \citenamefont {Mazza},\ and\ \citenamefont {Fazio}}]{BiellaFazio2016}%
  \BibitemOpen
  \bibfield  {author} {\bibinfo {author} {\bibfnamefont {A.}~\bibnamefont
  {Biella}}, \bibinfo {author} {\bibfnamefont {A.}~\bibnamefont {De~Luca}},
  \bibinfo {author} {\bibfnamefont {J.}~\bibnamefont {Viti}}, \bibinfo {author}
  {\bibfnamefont {D.}~\bibnamefont {Rossini}}, \bibinfo {author} {\bibfnamefont
  {L.}~\bibnamefont {Mazza}},\ and\ \bibinfo {author} {\bibfnamefont
  {R.}~\bibnamefont {Fazio}},\ }\bibfield  {title} {\bibinfo {title} {Energy
  transport between two integrable spin chains},\ }\href
  {https://doi.org/10.1103/PhysRevB.93.205121} {\bibfield  {journal} {\bibinfo
  {journal} {Phys. Rev. B}\ }\textbf {\bibinfo {volume} {93}},\ \bibinfo
  {pages} {205121} (\bibinfo {year} {2016})}\BibitemShut {NoStop}%
\bibitem [{\citenamefont {Mascarenhas}\ \emph {et~al.}(2017)\citenamefont
  {Mascarenhas}, \citenamefont {Giudice},\ and\ \citenamefont
  {Savona}}]{MascarenhasSavona2017}%
  \BibitemOpen
  \bibfield  {author} {\bibinfo {author} {\bibfnamefont {E.}~\bibnamefont
  {Mascarenhas}}, \bibinfo {author} {\bibfnamefont {G.}~\bibnamefont
  {Giudice}},\ and\ \bibinfo {author} {\bibfnamefont {V.}~\bibnamefont
  {Savona}},\ }\bibfield  {title} {\bibinfo {title} {A nonequilibrium quantum
  phase transition in strongly coupled spin chains},\ }\href
  {https://doi.org/10.22331/q-2017-12-20-40} {\bibfield  {journal} {\bibinfo
  {journal} {Quantum}\ }\textbf {\bibinfo {volume} {1}},\ \bibinfo {pages} {40}
  (\bibinfo {year} {2017})}\BibitemShut {NoStop}%
\bibitem [{\citenamefont {Biella}\ \emph {et~al.}(2019)\citenamefont {Biella},
  \citenamefont {Collura}, \citenamefont {Rossini}, \citenamefont {De~Luca},\
  and\ \citenamefont {Mazza}}]{BiellaMazza2019}%
  \BibitemOpen
  \bibfield  {author} {\bibinfo {author} {\bibfnamefont {A.}~\bibnamefont
  {Biella}}, \bibinfo {author} {\bibfnamefont {M.}~\bibnamefont {Collura}},
  \bibinfo {author} {\bibfnamefont {D.}~\bibnamefont {Rossini}}, \bibinfo
  {author} {\bibfnamefont {A.}~\bibnamefont {De~Luca}},\ and\ \bibinfo {author}
  {\bibfnamefont {L.}~\bibnamefont {Mazza}},\ }\bibfield  {title} {\bibinfo
  {title} {Ballistic transport and boundary resistances in inhomogeneous
  quantum spin chains},\ }\href {https://doi.org/10.1038/s41467-019-12784-4}
  {\bibfield  {journal} {\bibinfo  {journal} {Nat Commun}\ }\textbf {\bibinfo
  {volume} {10}},\ \bibinfo {pages} {4820} (\bibinfo {year}
  {2019})}\BibitemShut {NoStop}%
\bibitem [{\citenamefont {Ljubotina}\ \emph {et~al.}(2017)\citenamefont
  {Ljubotina}, \citenamefont {{\v Z}nidari{\v c}},\ and\ \citenamefont
  {Prosen}}]{LjubotinaProsen2017}%
  \BibitemOpen
  \bibfield  {author} {\bibinfo {author} {\bibfnamefont {M.}~\bibnamefont
  {Ljubotina}}, \bibinfo {author} {\bibfnamefont {M.}~\bibnamefont {{\v
  Z}nidari{\v c}}},\ and\ \bibinfo {author} {\bibfnamefont {T.}~\bibnamefont
  {Prosen}},\ }\bibfield  {title} {\bibinfo {title} {Spin diffusion from an
  inhomogeneous quench in an integrable system},\ }\href
  {https://doi.org/10.1038/ncomms16117} {\bibfield  {journal} {\bibinfo
  {journal} {Nat Commun}\ }\textbf {\bibinfo {volume} {8}},\ \bibinfo {pages}
  {16117} (\bibinfo {year} {2017})}\BibitemShut {NoStop}%
\bibitem [{\citenamefont {{\v Z}nidari{\v c}}\ and\ \citenamefont
  {Ljubotina}(2018)}]{ZnidaricLjubotina2018}%
  \BibitemOpen
  \bibfield  {author} {\bibinfo {author} {\bibfnamefont {M.}~\bibnamefont {{\v
  Z}nidari{\v c}}}\ and\ \bibinfo {author} {\bibfnamefont {M.}~\bibnamefont
  {Ljubotina}},\ }\bibfield  {title} {\bibinfo {title} {Interaction instability
  of localization in quasiperiodic systems},\ }\href
  {https://doi.org/10.1073/pnas.1800589115} {\bibfield  {journal} {\bibinfo
  {journal} {Proc. Natl. Acad. Sci. USA}\ }\textbf {\bibinfo {volume} {115}},\
  \bibinfo {pages} {4595} (\bibinfo {year} {2018})}\BibitemShut {NoStop}%
\bibitem [{\citenamefont {Ljubotina}\ \emph {et~al.}(2019)\citenamefont
  {Ljubotina}, \citenamefont {{\v Z}nidari{\v c}},\ and\ \citenamefont
  {Prosen}}]{LjubotinaProsen2019}%
  \BibitemOpen
  \bibfield  {author} {\bibinfo {author} {\bibfnamefont {M.}~\bibnamefont
  {Ljubotina}}, \bibinfo {author} {\bibfnamefont {M.}~\bibnamefont {{\v
  Z}nidari{\v c}}},\ and\ \bibinfo {author} {\bibfnamefont {T.}~\bibnamefont
  {Prosen}},\ }\bibfield  {title} {\bibinfo {title} {Kardar-{{Parisi}}-{{Zhang
  Physics}} in the {{Quantum Heisenberg Magnet}}},\ }\href
  {https://doi.org/10.1103/PhysRevLett.122.210602} {\bibfield  {journal}
  {\bibinfo  {journal} {Phys. Rev. Lett.}\ }\textbf {\bibinfo {volume} {122}},\
  \bibinfo {pages} {210602} (\bibinfo {year} {2019})}\BibitemShut {NoStop}%
\bibitem [{\citenamefont {Balachandran}\ \emph {et~al.}(2018)\citenamefont
  {Balachandran}, \citenamefont {Benenti}, \citenamefont {Pereira},
  \citenamefont {Casati},\ and\ \citenamefont
  {Poletti}}]{BalachandranPoletti2018b}%
  \BibitemOpen
  \bibfield  {author} {\bibinfo {author} {\bibfnamefont {V.}~\bibnamefont
  {Balachandran}}, \bibinfo {author} {\bibfnamefont {G.}~\bibnamefont
  {Benenti}}, \bibinfo {author} {\bibfnamefont {E.}~\bibnamefont {Pereira}},
  \bibinfo {author} {\bibfnamefont {G.}~\bibnamefont {Casati}},\ and\ \bibinfo
  {author} {\bibfnamefont {D.}~\bibnamefont {Poletti}},\ }\bibfield  {title}
  {\bibinfo {title} {Perfect {{Diode}} in {{Quantum Spin Chains}}},\ }\href
  {https://doi.org/10.1103/PhysRevLett.120.200603} {\bibfield  {journal}
  {\bibinfo  {journal} {Phys. Rev. Lett.}\ }\textbf {\bibinfo {volume} {120}},\
  \bibinfo {pages} {200603} (\bibinfo {year} {2018})}\BibitemShut {NoStop}%
\bibitem [{\citenamefont {Bushong}\ \emph {et~al.}(2005)\citenamefont
  {Bushong}, \citenamefont {Sai},\ and\ \citenamefont
  {Di~Ventra}}]{BushongDiVentra2005}%
  \BibitemOpen
  \bibfield  {author} {\bibinfo {author} {\bibfnamefont {N.}~\bibnamefont
  {Bushong}}, \bibinfo {author} {\bibfnamefont {N.}~\bibnamefont {Sai}},\ and\
  \bibinfo {author} {\bibfnamefont {M.}~\bibnamefont {Di~Ventra}},\ }\bibfield
  {title} {\bibinfo {title} {Approach to {{Steady}}-{{State Transport}} in
  {{Nanoscale Conductors}}},\ }\href {https://doi.org/10.1021/nl0520157}
  {\bibfield  {journal} {\bibinfo  {journal} {Nano Lett.}\ }\textbf {\bibinfo
  {volume} {5}},\ \bibinfo {pages} {2569} (\bibinfo {year} {2005})}\BibitemShut
  {NoStop}%
\bibitem [{\citenamefont {Karrasch}\ \emph {et~al.}(2013)\citenamefont
  {Karrasch}, \citenamefont {Ilan},\ and\ \citenamefont
  {Moore}}]{KarraschMoore2013}%
  \BibitemOpen
  \bibfield  {author} {\bibinfo {author} {\bibfnamefont {C.}~\bibnamefont
  {Karrasch}}, \bibinfo {author} {\bibfnamefont {R.}~\bibnamefont {Ilan}},\
  and\ \bibinfo {author} {\bibfnamefont {J.~E.}\ \bibnamefont {Moore}},\
  }\bibfield  {title} {\bibinfo {title} {Nonequilibrium thermal transport and
  its relation to linear response},\ }\href
  {https://doi.org/10.1103/PhysRevB.88.195129} {\bibfield  {journal} {\bibinfo
  {journal} {Phys. Rev. B}\ }\textbf {\bibinfo {volume} {88}},\ \bibinfo
  {pages} {195129} (\bibinfo {year} {2013})}\BibitemShut {NoStop}%
\bibitem [{\citenamefont {Brenes}\ \emph {et~al.}(2020)\citenamefont {Brenes},
  \citenamefont {LeBlond}, \citenamefont {Goold},\ and\ \citenamefont
  {Rigol}}]{BrenesRigol2020}%
  \BibitemOpen
  \bibfield  {author} {\bibinfo {author} {\bibfnamefont {M.}~\bibnamefont
  {Brenes}}, \bibinfo {author} {\bibfnamefont {T.}~\bibnamefont {LeBlond}},
  \bibinfo {author} {\bibfnamefont {J.}~\bibnamefont {Goold}},\ and\ \bibinfo
  {author} {\bibfnamefont {M.}~\bibnamefont {Rigol}},\ }\bibfield  {title}
  {\bibinfo {title} {Eigenstate {{Thermalization}} in a {{Locally Perturbed
  Integrable System}}},\ }\href
  {https://doi.org/10.1103/PhysRevLett.125.070605} {\bibfield  {journal}
  {\bibinfo  {journal} {Phys. Rev. Lett.}\ }\textbf {\bibinfo {volume} {125}},\
  \bibinfo {pages} {070605} (\bibinfo {year} {2020})}\BibitemShut {NoStop}%
\bibitem [{\citenamefont {LeBlond}\ and\ \citenamefont
  {Rigol}(2020)}]{LeBlondRigol2020}%
  \BibitemOpen
  \bibfield  {author} {\bibinfo {author} {\bibfnamefont {T.}~\bibnamefont
  {LeBlond}}\ and\ \bibinfo {author} {\bibfnamefont {M.}~\bibnamefont
  {Rigol}},\ }\bibfield  {title} {\bibinfo {title} {Eigenstate thermalization
  for observables that break {{Hamiltonian}} symmetries and its counterpart in
  interacting integrable systems},\ }\href
  {https://doi.org/10.1103/PhysRevE.102.062113} {\bibfield  {journal} {\bibinfo
   {journal} {Phys. Rev. E}\ }\textbf {\bibinfo {volume} {102}},\ \bibinfo
  {pages} {062113} (\bibinfo {year} {2020})}\BibitemShut {NoStop}%
\bibitem [{\citenamefont {{\L}yd{\.z}ba}\ \emph {et~al.}(2021)\citenamefont
  {{\L}yd{\.z}ba}, \citenamefont {Zhang}, \citenamefont {Rigol},\ and\
  \citenamefont {Vidmar}}]{LydzbaVidmar2021}%
  \BibitemOpen
  \bibfield  {author} {\bibinfo {author} {\bibfnamefont {P.}~\bibnamefont
  {{\L}yd{\.z}ba}}, \bibinfo {author} {\bibfnamefont {Y.}~\bibnamefont
  {Zhang}}, \bibinfo {author} {\bibfnamefont {M.}~\bibnamefont {Rigol}},\ and\
  \bibinfo {author} {\bibfnamefont {L.}~\bibnamefont {Vidmar}},\ }\bibfield
  {title} {\bibinfo {title} {Single-particle eigenstate thermalization in
  quantum-chaotic quadratic {{Hamiltonians}}},\ }\href@noop {} {\bibfield
  {journal} {\bibinfo  {journal} {arXiv:2109.06895}\ } (\bibinfo {year}
  {2021})}\BibitemShut {NoStop}%
\bibitem [{NSC()}]{NSCC}%
  \BibitemOpen
  \href@noop {} {}\bibinfo {howpublished} {https://www.nscc.sg/}\BibitemShut
  {NoStop}%
\end{thebibliography}
\end{document}


\title{Supplementary Materials: \\Typicality of nonequilibrium (quasi-)steady currents}               

\author{Xiansong Xu} 
\affiliation{\sutd}
\author{Chu Guo} 
\affiliation{Henan Key Laboratory of Quantum Information and Cryptography, Zhengzhou,
Henan 450000, China}
\affiliation{\hnu}
\author{Dario Poletti}
\affiliation{\sutd} 
\affiliation{\sutdepd}

\date{\today}
\maketitle
\tableofcontents

\section{Perturbative and exact currents}
In this section we derive the perturbative current expression (Eq. (5) of the main paper) similar to Ref.~\cite{ZhouZhang2020} which focused on transport problems through a weakly coupled interface. In this work, we consider two finite quantum spin chains with a total length $N$,
\begin{align}
    H & = \HL + \HR + V, \\
    V &= \gamma \BL \otimes \BR, \label{eq:overall_Hamiltonian}
\end{align}
where $\BL$ and $\BR$ are operators acting respectively on the left and right baths and $\gamma$ is the coupling strength. We work in units for which $\hbar=1$. 

Since the total energy is a conserved quantity for each region, the energy current operators can be defined as the rate of change of the Hamiltonian from the left region given by 
\begin{align}
    I^{\rm L} &=-\frac{d \HL}{dt}=-\im \gamma \left[\BL,\HL\right] \otimes \BR =  \dBL \otimes \BR,
    \label{eq: currentop}
\end{align}
where we defined $\dBL \equiv -\im \gamma \left[\BL,\HL\right]$.

To find the current $\cur^{\rm L}=\tr\left( \rho I^{\rm L} \right)$, we consider the Dyson series expansion of the total density operator in the interaction picture $\rhoI(t) = \expe{\im (\HL + \HR) t} \rho \expe{-\im (\HL + \HR) t}$ of the composite system $H$ with respect to the coupling strength $\gamma$,
\begin{align}
    \rhoI(t) = \rhoI(0)- \im \int_{0}^{t} d \tau\left[\VI\left(\tau\right),
    \rhoI(0)\right]+O\left(\VI^{2}\right).
    \label{eq: dyson}
\end{align}
The corresponding current operator in interaction picture is thus $I^{\rm L}(t) = \expe{\im (\HL + \HR) t} I^{\rm L} \expe{-\im (\HL + \HR) t}$ and $\cur^{\rm L} =  \tr\left( \tilde{\rho}(t) I^{\rm L}(t) \right)$.
By considering a separable initial condition where the initial density operator is the tensor product of two operators acting only on the left of right bath, $\rho(0) = \rho_{\rm L}(0) \bigotimes \rho_{\rm R}(0)$, it follows from Eqs. (\ref{eq: currentop}, \ref{eq: dyson}) that a perturbative current expression
can be found as 
\begin{align}
    \cur^{\rm L}(t)=&\mathcal{\dot{B}}^{\rm L}(t) \mathcal{B}^{\rm R}(t) - \im \int_{0}^{t} d \tau \left[
    \overrightarrow{\mathcal{C}}_{\rm L}(t, \tau) {\mathcal{C}}_{\rm R}(t, \tau)
    -\overleftarrow{\mathcal{C}}_{\rm L}(\tau,t) {\mathcal{C}}_{\rm R}(\tau,t) \right] 
    \label{eq: current_general}
\end{align}
where $\mathcal{\dot{B}}^{\rm L}(t)=\tr_{\rm L}\left[\rho_{\rm L} \dBL(t) \right]$ and
$\mathcal{B}^{\rm R}(t)=\tr_{\rm R}\left[\rho_{\rm R} \BR(t) \right]$ with $\dBL(t) = \expe{\im \HL t} \dBL \expe{-\im \HL t}$ and $\BR(t) = \expe{\im \HR t} \BR \expe{-\im \HR t}$. Here $\mathcal{C}^{\rm L}$ and
$\mathcal{C}^{\rm R}$ are the correlation functions defined as
\begin{align}
    \overrightarrow{\mathcal{C}}_{\rm L}(t, \tau) &=\tr_{\rm L}\left[\dBL(t) \BL(\tau) \rho_{\rm L }(0) \right], \label{eq: correlationS1} \\
    \overleftarrow{\mathcal{C}}_{\rm L}(\tau, t) &=\tr_{\rm L}\left[\BL(\tau) \dBL(t) \rho_{\rm L }(0)  \right], \label{eq: correlationS2} \\
    {\mathcal{C}}_{\rm R}(t, \tau) &=\tr_{\rm R}\left[\BR(t) \BR(\tau) \rho_{\rm R}(0) \right].
    \label{eq: correlationS3}
\end{align}
Up to this point, we have not yet specified the density operator $\rho_{\rm L}$ and $\rho_{\rm R}$. When the density operators are such that $\left[ \HL, \rho_{\rm L} \right] = \left[ \HR, \rho_{\rm R} \right] = 0$, e.g., when the density operators are given by some equilibrium states, the correlation functions can be simplified in a time translational invariant form (in the sense that it only depends on the difference of $t$ and $\tau$)  
\begin{align}
    \overrightarrow{\mathcal{C}}_{\rm L}(t - \tau) &=\tr_{\rm L}\left[\dBL(t - \tau) \BL \rho_{\rm L }(0) \right], \label{eq: lcorr_eq1}\\
    \overleftarrow{\mathcal{C}}_{\rm L}(\tau - t) &=\tr_{\rm L}\left[\BL(\tau - t) \dBL \rho_{\rm L }(0)  \right], \label{eq: lcorr_eq2}\\
    {\mathcal{C}}_{\rm R}(t - \tau) &=\tr_{\rm R}\left[\BR(t - \tau) \BR \rho_{\rm R}(0) \right].
    \label{eq: rcorr_eq}
\end{align}
One can further simplify the current expression by considering a change of variable $t-\tau \rightarrow \tau'$ and the current can then be expressed as 
\begin{align}
    \cur^{\rm L}(t)= - \im \int_{-t}^{t} d \tau^\prime
    \overrightarrow{\mathcal{C}}_{\rm L}(\tau^\prime) \mathcal{C}_{\rm R}(\tau^\prime).
    \label{eq: current_eq}
\end{align}
because $\overrightarrow{\mathcal{C}}_{\rm L}(\tau^\prime) = -\overleftarrow{\mathcal{C}}_{\rm L}(\tau^\prime)$ and $\mathcal{\dot{B}}^{\rm L}(t) \mathcal{B}^{\rm R}(t)=0$ for equilibrium states. 

The first remark is that the above correlation functions Eqs. (\ref{eq: lcorr_eq1}, \ref{eq: lcorr_eq2}, \ref{eq: rcorr_eq}) and the current expression Eq. (\ref{eq: current_eq}) only apply to equilibrium states such as eigenstates, microcanonical distributions, and canonical distributions and they do not apply to typical initial states. Typical initial states in general do not commute with the Hamiltonian $\HL$ or $\HR$. As a consequence, the time translational invariance condition does not hold and the free evolution term $\mathcal{\dot{B}}^{\rm L}(t) \mathcal{B}^{\rm R}(t)$ does not vanish. Thus, in Fig. 1 in the main paper, the perturbative results for typical initial conditions are obtained via Eq. (\ref{eq: current_general}) and those for equilibrium states are obtained via Eq. (\ref{eq: current_eq}). The second remark is that these perturbative current expressions assume that the environments are not affected by the interaction, in a similar manner commonly used in perturbative quantum master equations such as Born-Markov master equation \cite{BreuerPetruccione2007}. 

For numerical exact calculations to measure the current, we follow the current operator definition given by Eq. (\ref{eq: currentop}) and the full expression of the current operator $I^{\rm L}$, with respect to the left environment Hamiltonian in Eq. (1) of the main paper, and here in Eq. (\ref{eq: hamiltonianS}), is given by 
\begin{align}
    I^{\rm L} = &  -\frac{d \HL}{dt} = -\im \left[ V, \HL \right] =  -\im \gamma \left[ \sx{N_{\rm L}}\sx{N_{\rm L}+1} , \HL  \right] \nonumber \\
    = & -2\gamma \left( J_{zz} \sz{N_{\rm L}-1} \sy{N_{\rm L}} \sx{N_{\rm L}+1} + J_{yz} \sy{N_{\rm L}-1} \sy{N_{\rm L}} \sx{N_{\rm L}+1} + h_z \sy{N_{\rm L}}\sx{N_{\rm L}+1} \right)
\end{align}
Correspondingly, the current operator for right bath can be defined as the energy entering the right part, that is     
\begin{align}
    I^{\rm R} = &  \frac{d \HR}{dt} = \im \left[ V, \HR \right] =  \im \gamma \left[ \sx{N_{\rm L}}\sx{N_{\rm L}+1} , \HR \right] \nonumber \\
    = & 2\gamma \left( J_{zz} \sx{N_{\rm L}} \sy{N_{\rm L}+1} \sz{N_{\rm L}+2} - J_{yz} \sx{N_{\rm L}} \sz{N_{\rm L}+1} \sz{N_{\rm L}+2} + h_z \sx{N_{\rm L}}\sy{N_{\rm L}+1} \right) \label{eq:current_right}
\end{align}

\section{Dynamical typicality and typical correlation functions}
The perturbative current expression given by Eq. (\ref{eq: current_eq}) allows us to extend the notion of typicality to currents via the correlation functions. We first show the typicality of the correlation functions in Eqs. (\ref{eq: correlationS1}, \ref{eq: correlationS2}, \ref{eq: correlationS3}) in a similar manner for isolated systems with respect to some typical pure states $\rhotyp$ drawn from a given $\rhomic$. If follows that one can show that the typical behavior will be inherited by the perturbative expression for the currents given in Eq. (\ref{eq: current_general}).

More in detail, if dynamical typicality holds, then one can write that for vast majority of the typical initial states,
\begin{align}
    \overrightarrow{\mathcal{C}}_{\rm L}^{\rm typ}  \approx \overrightarrow{\mathcal{C}}_{\rm L}^{\rm mic}, \quad \overleftarrow{\mathcal{C}}_{\rm L}^{\rm typ}  \approx \overleftarrow{\mathcal{C}}_{\rm L}^{\rm mic}, \quad {\mathcal{C}}_{\rm R}^{\rm typ}  \approx {\mathcal{C}}_{\rm R}^{\rm mic},
\end{align}
where the superscripts refer to typical-state initial condition and microcanonical initial condition. One can examine the expectation value of these typical correlation functions and it can be shown that they are equivalent to the microcanonical ones, i.e., $\Exp \left[\mathcal{C}^{\rm typ}\right] = \mathcal{C}^{\rm mic}$. For example, considering $\overrightarrow{\mathcal{C}}_{\rm L}^{\rm typ}(t,\tau)$ given by Eq. (\ref{eq: correlationS1}) and some typical initial states, $\left(\rho_{\rm L}^{\rm typ}\right)_{ij}= c_i^* c_j$ defined in Eq.~(3) of the main paper,
\begin{align}
    \overrightarrow{\mathcal{C}}_{\rm L}^{\rm typ} (t, \tau)
    =& \tr_{\rm L}\left[ \dBL(t) \BL(\tau)  \rho_{\rm L}^{\rm typ}\right] 
    = -\im \gamma \sum_{ij}^{d_{\rm L}} \sum_{k}^{D} c_i^* c_j  \expe{\im (\Delta_{ik} t +  \Delta_{kj}\tau )} \Delta_{ki} \BL_{ik} \BL_{kj}, 
\end{align}
with $\Delta_{ij} = E_i - E_j$. 
The expectation value of $\overrightarrow{\mathcal{C}}_{\rm L}^{\rm typ} (t, \tau)$ with respect to the ensemble of typical states is  
\begin{align}
    \Exp\left[ \overrightarrow{\mathcal{C}}_{\rm L}^{\rm typ} (t,\tau)\right]
    = & -\im \gamma \sum^D_k \sum_{ij}^{d_{\rm L}} \Exp{\left[c_i^* c_j\right]}  \expe{\im (\Delta_{ik}t +\Delta_{kj}\tau)} \Delta_{ki} \BL_{ik} \BL_{kj} 
    =  -\im \gamma\frac{1}{d_{\rm L}} \sum^D_k \sum_{i}^{d_{\rm L}} \expe{\im \Delta_{ik} (t-\tau)} \Delta_{ki} \BL_{ik} \BL_{ki} 
    =  \overrightarrow{\mathcal{C}}_{\rm L}^{\rm mic}(t,\tau),
\end{align}
where $\Exp\left[c_i^*c_j\right]=\delta_{i,j}/d_{\rm L}$. Similar conclusions hold for $\overleftarrow{\mathcal{C}}_{\rm L}^{\rm typ}$ and $\mathcal{C}_{\rm R}^{\rm typ}$. One can thus infer that the typicality for currents holds, at least in the weak coupling limit for which perturbative expressions are accurate, as the correlation functions for left and right environments are independent to each other. Hence, one can expect that
\begin{align}
    \Exp\left[\cur^{\rm L}_{\rm typ}(t)\right]=&\Exp\left[\mathcal{\dot{B}}^{\rm L}_{\rm typ}(t) \mathcal{B}^{\rm R}_{\rm typ}(t)\right] - \im \int_{0}^{t} d \tau
    \Exp\left[\overrightarrow{\mathcal{C}}_{\rm L}^{\rm typ}(t,\tau) \mathcal{C}_{\rm R}^{\rm typ}(t,\tau) - \overleftarrow{\mathcal{C}}_{\rm L}^{\rm typ}(\tau,t) \mathcal{C}_{\rm R}^{\rm typ}(\tau,t)\right]  \\
    =& - \im \int_{0}^{t} d \tau 
    \left[\overrightarrow{\mathcal{C}}_{\rm L}^{\rm mic}(t,\tau) \mathcal{C}^{\rm mic}_{\rm R}(t,\tau)
          -\overleftarrow{\mathcal{C}}_{\rm L}^{\rm mic}(\tau,t) \mathcal{C}^{\rm mic}_{\rm R}(\tau, t)
    \right]
     = \cur^{\rm L}_{\rm mic}(t),
\end{align}
where $\Exp\left[\mathcal{\dot{B}}^{\rm L}_{\rm typ}(t) \mathcal{B}^{\rm R}_{\rm typ}(t)\right]= \Exp\left[\mathcal{\dot{B}}^{\rm L}_{\rm typ}(t)\right]\Exp\left[\mathcal{B}^{\rm R}_{\rm typ}(t)\right]$ vanishes as
\begin{align}
    \Exp\left[\mathcal{\dot{B}}^{\rm L}_{\rm typ}(t)\right] = \Exp\left[\sum_{ij}^{d_{\rm L}} c_i^*c_j \Delta_{ji}B_{ij}\right] = 0.
\end{align}
Furthermore, as shown in Fig. 2 of the main paper, and as can be seen in Fig. \ref{fig: sstvsL}, we have checked numerically that the variance of the currents is bounded and decreases with the system size. 

\section{Current expression via eigenstate thermalization hypothesis}

\subsection{Correlation functions via eigenstate thermalization hypothesis}
Since each of the environment are nonintegrable and thus follows the eigenstate thermalization hypothesis, one can evaluate the correlation functions $\overrightarrow{\mathcal{C}}_{\rm L}(\tau)$, $\overleftarrow{\mathcal{C}}_{\rm L}(\tau)$, and $\mathcal{C}_{\rm R}(\tau)$ in a manner similar to Refs. \cite{LuitzBarLev2016, DAlessioRigol2016, Srednicki1996} and thus a perturbative current expression. Considering the ETH ansatz for the operator $B$, 
\begin{align}
	B_{ij} &= \Braket{B}(E) \delta_{ij} + \expe{-\frac{S(E)}{2}} f(E,\omega) R_{ij} \label{eq:ETH_ansatz}
\end{align}
where $B_{ij}$ are the elements of $B$ in the energy basis, $E_i$ an energy eigenvalue, $E = (E_i + E_j)/2$, $\omega = E_j - E_i$ and $R_{ij}$ is a matrix whose elements follow a normal distribution with zero mean and unit variance.
We consider the correlation function for a single eigenstates with eigenvalue $E_i$ $\overrightarrow{\mathcal{C}}_{\rm L}^{\rm eig}(\tau)$ which, using Eq. (\ref{eq: lcorr_eq1}), can be written as 
\begin{align}
	\overrightarrow{\mathcal{C}}_{\rm L}^{\rm eig}(\tau) = & -\im \gamma \sum^D_j \expe{-i \omega \tau} \omega \BL_{ij} \BL_{ji}.
\end{align}
It follows that the corresponding correlation expression via ETH is  
\begin{align}
    \overrightarrow{\mathcal{C}}_{\rm L}^{\rm ETH}(\tau) \approx & -\im \gamma \sum^D_j \omega \expe{-\im \omega \tau } \expe{-S(E)} |f_{\rm L}(E, \omega)|^2  |R_{ij}|^2 .  
\end{align}
By taking the continuum limit where $\sum^{D}_{j} \rightarrow \int dE_j \; \expe{S(E_j)} = \int d\omega \; \expe{S(E_i + \omega)}$, we have 
\begin{align}
  \overrightarrow{\mathcal{C}}_{\rm L}^{\rm ETH}(\tau)  
    \approx & -\im \gamma \int dE_j \; \dfrac{\expe{S(E_i + \omega)}}{\expe{S(E_i + \omega/2)}} \; \omega \;  
    |f_{\rm L}(E_i + \frac{\omega}{2}, \omega)|^2 |R_{ij}|^2 \expe{-\im \omega \tau} \nonumber \\
    \approx & -\im \gamma  \int d\omega \; G(E_i, \omega) \; \omega \; 
    |f_{\rm L}(E_i + \frac{\omega}{2}, \omega)|^2 \expe{-\im \omega \tau} \label{eq:CL_ETH1}. 
\end{align}
To obtain the last expression in Eq. (\ref{eq:CL_ETH1}) we have defined  
\begin{align}
    G(E_i, \omega) \equiv \dfrac{\expe{S(E_i + \omega)}}{\expe{S(E_i + \omega/2)}},
    \label{eq: dos_ratio}
\end{align}
and we have also set $|R_{ij}|^2=1$ because this is the value of the variance of the elements of the matrix $R_{ij}$. Finally, we can safely drop the $\omega/2$ shift in $f_{\rm L}$, and write $f_{\rm L}(E_i, \omega)$, as  $f_{\rm L}$ varies mildly with respect to $E_i$ and quickly decays with respect to $\omega$. This results in  
\begin{align}
  \overrightarrow{\mathcal{C}}_{\rm L}^{\rm ETH}(\tau) \approx & -\im \gamma  \int d\omega \; G(E_i, \omega) \; \omega \; |f_{\rm L}(E_i, \omega)|^2 \expe{-\im \omega \tau} \label{eq:CL_ETH}. 
\end{align}
Following the same steps, we can also write a simpler to evaluate expression for $\mathcal{C}_{\rm R}^{\rm ETH}(\tau)$, which is 
\begin{align}
    \mathcal{C}_{\rm R}^{\rm ETH}(\tau)  & \approx \int  d\omega \; G(E_i, \omega) \; \lvert f_{\rm R}(E_i, \omega) \rvert^2 \expe{-\im \omega \tau}.  \label{eq:CR_ETH}      
\end{align}

\subsection{Bath Hamiltonian and local operators properties: diagonal and off-diagonal}
\begin{figure}
    \centering
    \includegraphics[width=0.6\columnwidth]{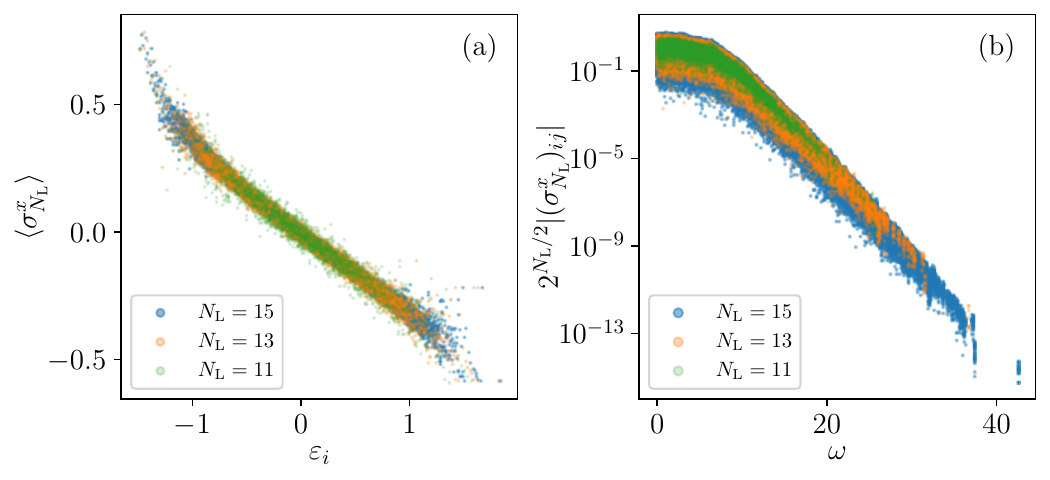}
    \caption{(a) $\Braket{\sx{N_{\rm L}}}$ versus the eigenenergy per spin $\varepsilon_i = E_i/N_{\rm L}$ for eigenstate $\Ket{i}$ and (b) Absolute value of the off-diagonal elements of $\sx{{N}_{\rm L}}$ scaled by $2^{{N}_{\rm L}/2}$ with respect to the corresponding level spacings $\omega=E_i - E_j$ for different system sizes $N_{\rm L}=15$ (blue), $N_{\rm L}=13$ (orange), and $N_{\rm L}=11$ (green). For panel (b), we consider a fixed energy per spin $\varepsilon_{\rm L} = 5/12$ with energy shell width per spin $\delta_{\rm L}=0.1/12$.} 
    \label{fig: sx}
\end{figure}

In this work we have chosen a nonintegrable local Hamiltonian which has no conserved quantities except for the energy. In the specific, although the results do not depend on this particular choice of Hamiltonian, we have considered the following one for the left bath 
\begin{align}
    \HL = \sum_{n=1}^{N_{\rm L}-1} \left( J_{zz} \sz{n}\sz{n+1} + J_{yz} \sy{n}\sz{n+1} \right)+ \sum^{N_{\rm L}}_{n=1} \left(h_x \sx{n} + h_z \sz{n} \right), 
    \label{eq: hamiltonianS}
\end{align}
and the same for the right bath's Hamiltonian $\HL$, except that the site number $n$ goes from $N_{\rm L}+ 1$ to $N_{\rm L}+ N_{\rm R}$. 
Dynamical typicality requires common initial expectation values for the observables of interest, a property that readily occurs in systems following ETH.  In Fig.~\ref{fig: sx}(a) we show $\langle i | \sx{N_{\rm L}} | i \rangle_{\rm L}$ versus the energy per spin for the eigenstate $|i\rangle_{\rm L}$ which, as the system size increases, follows a smooth curve as expected from ETH. 

For the off-diagonal elements, in Fig.~\ref{fig: sx}(b) we show the off-diagonal elements $\langle i | \sx{N_{\rm L}} | j \rangle_{\rm L}$ multiplied by the square root of the Hilbert space size $2^{N_{\rm L}/2}$, versus the energy difference $\omega$. To make this plot more relevant to the main paper, here we have considered a window of energies for the left bath (we can transfer these results to the right bath since it follows the same Hamiltonian) $\Xi^{\rm L} = N_{\rm L}[\varepsilon_{\rm L}-\delta_{\rm L}/2, \varepsilon_{\rm L}+\delta_{\rm L}/2]$ where $\varepsilon_{\rm L}$ is an average energy per particle, while $\delta_{\rm L}$ indicates the linear scaling factor of the width of the shell. Furthermore, the off-diagonal elements $\langle i | \sx{N_{\rm L}} | j \rangle_{\rm L}$ are selected in the energy window such that $|(E_i + E_j)/(2N_{\rm L}\varepsilon_{\rm L}) - 1| < N_{\rm L}\delta_{\rm L}$ \cite{KhatamiRigol2013, JansenHeidrich-Meisner2019}. Also in Fig.~\ref{fig: sx}(b) we see the expected result from a system which follows ETH, i.e. that the (rescaled) off-diagonal elements tend to a smooth curve $f_{\rm L}(E_i, \omega)$. 

\subsection{Scaling analysis}
\begin{figure}
    \centering
    \includegraphics[width=\columnwidth]{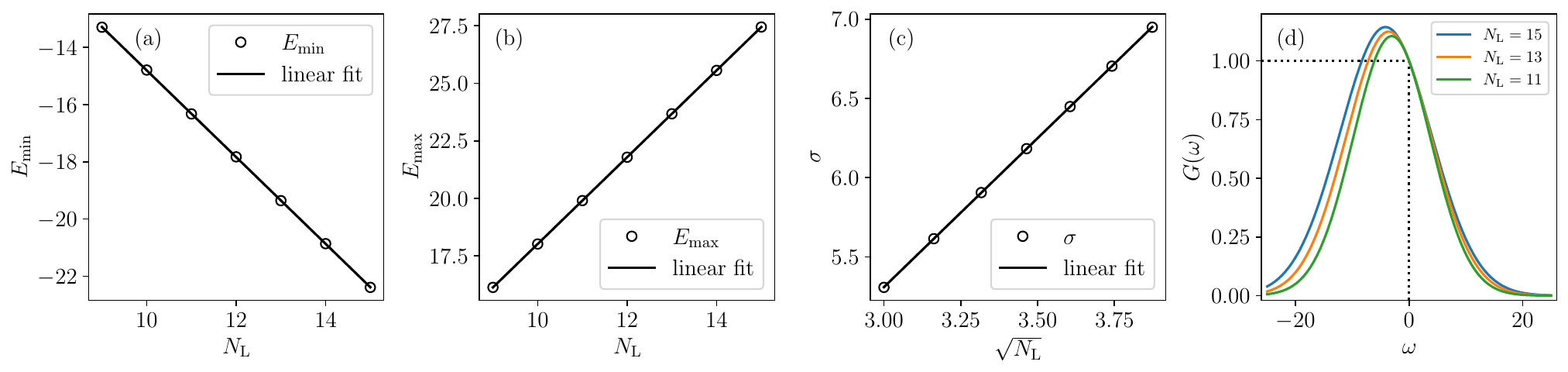}
    \caption{(a) Maximum energy and (b) the ground state energy with system size $N$ for a single system. (c) The standard deviation of the Gaussian-like density of states with $\sqrt{N}$. (d) The ratio of shifted density of states $G(\omega)$ versus $\omega$ for different system sizes $N_{\rm L} = 11$ (green), $13$ (orange), $15$ (blue).}
    \label{fig: scaling}
\end{figure}

In order to evaluate Eqs.~(\ref{eq:CL_ETH}, \ref{eq:CR_ETH}) we need to further analyze the scaling of the different quantities included. The energy windows $\Xi_{\rm L/R}$ considered shift both in average value and in width linearly with the system size. This is due to the extensive nature of the Hamiltonians $\HL$ and $\HR$ used. One natural numerical check that one can do is to verify the scaling of the maximum and minimum value of the energy, respectively $E_{\rm max}$ and $E_{\rm min}$. These quantities are shown in Fig. \ref{fig: scaling} (a, b), which clearly depicts a linear scaling. In Fig. \ref{fig: scaling} (c), after fitting the many-body density of states $\sum_i\delta(E-E_i)$ with a Gaussian function, we show that the standard deviation of such a Gaussian-like density of states scales as a function $\sqrt{N_{\rm L}}$, which confirms that for this local Hamiltonian, the density of states indeed scales as the square root of the system size. This allows us to find the ratio $G(E_i,\omega) = \exp{[S(E_i + \omega)]} / \exp{[S(E_i+\omega/2)]}$ defined in Eq. (\ref{eq: dos_ratio}), which is depicted in Fig. \ref{fig: scaling}(d) for different sizes. Note that there exists a fixed point for $\omega = 0$ and the lines converge towards a single line near it.  

\begin{figure}
    \centering
    \includegraphics[width=0.6\columnwidth]{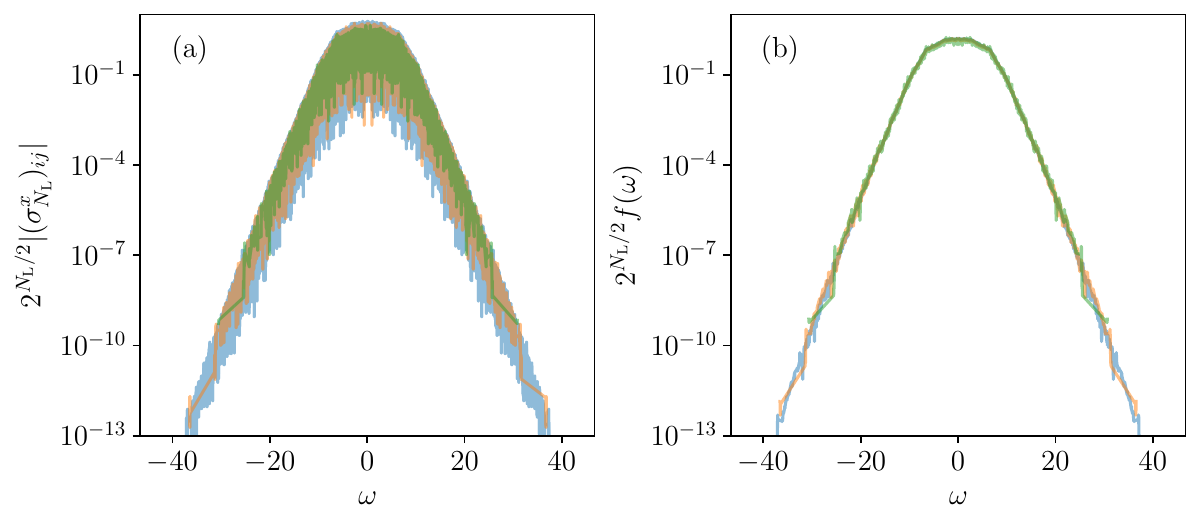}
    \caption{(a) Scaled off-diagonals elements with respect to the energy difference $\omega$ and (b) their moving averages for different system sizes $N_{\rm L}=11$ (green), $13$ (orange), $15$ (blue). The energy per spin is $\varepsilon_{\rm L}=5/12$ and $\delta_{\rm L} = 0.1/12$. The moving averages are computed for energy difference window $\delta \omega = 0.3/E^{\rm L} = 0.3/(N_{\rm L} \varepsilon_{\rm L})$.}
    \label{fig: offdiag-scaling}
\end{figure}

In Fig. \ref{fig: offdiag-scaling}, we show the function $f(E,\omega)$ for finite size systems with fixed energy per spin. The function $f(E,\omega)$ is extracted by considering the moving average of the off-diagonal elements. Fig. \ref{fig: offdiag-scaling}(a) shows the off-diagonal elements which are the same as Fig. \ref{fig: sx}(b) with the full range of $\omega$. Fig. \ref{fig: offdiag-scaling}(b) is the moving averages of the off-diagonal elements of panel (a) and we interpolate them as the envelope function $f(E, \omega)$ for a fixed energy per spin. The moving average is computed by considering averaged absolute values of the off-diagonal elements in a window $\delta\omega/E_{\rm L}$ \cite{JansenHeidrich-Meisner2019}. Different system sizes gives a consistent $f(E,\omega)$ function profile upon a scaling with $2^{N_{\rm L}/2}$ given by the prefactor in the ETH ansatz in Eq. (\ref{eq:ETH_ansatz}). 

An important consequence from the function $f$ profile, as demonstrated in Fig. \ref{fig: offdiag-scaling} (a, b), is that the contribution near $\omega = 0$ is the most significant one. This means that the fixed point in Fig. \ref{fig: scaling} contributes most significantly to the many-body density of states ratio and thus the correlation integral is also invariant under a scaling with respect to system size $N$.

\section{Equivalence of currents for left and right baths}
\begin{figure}[h!]
    \centering
    \includegraphics[width=0.5\columnwidth]{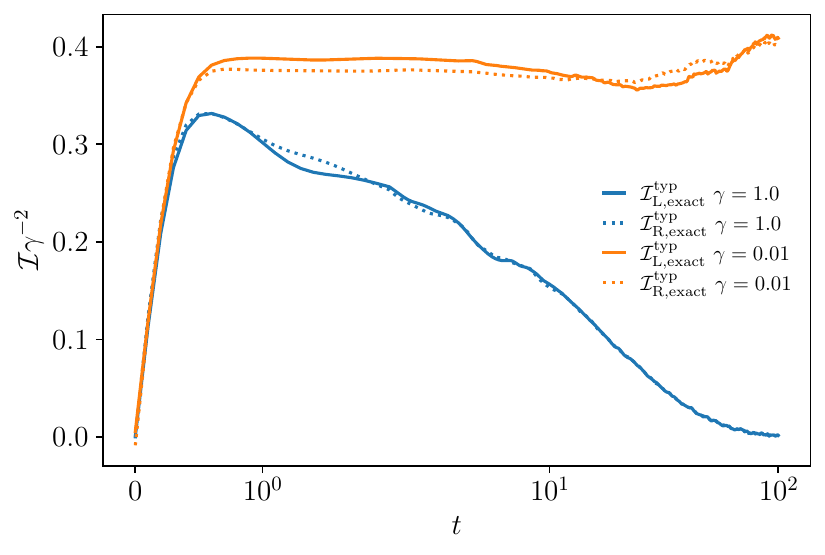}
    \caption{Exact typical currents from the left bath $\cur^{\rm exa}_{\rm L}$ (solid lines) and that entering the right bath the $\cur^{\rm exa}_{\rm R}$ (dotted lines) for strong couplings $\gamma=1.0$ (blue) and weak couplings $\gamma=0.01$ (orange). $N=N_{\rm L} +N_{\rm R} = 24$, $N_{\rm L} = N_{\rm R}=12$, $\delta_{\rm L} = \delta_{\rm R}=0.1/12$ and $\varepsilon_{\rm L} = -\varepsilon_{\rm R} = 5/12$ in units such that $J_{zz}=1$.}\label{fig:left_right_currents}
\end{figure}

In the main paper, we show the emergence of long-lasting steady currents, however, the data we show refers solely to the energy current leaving the left bath. For completeness, here we show that the current entering the right region, see Eq.~(\ref{eq:current_right}), coincides with the one leaving the right bath (up to small deviations due to energy being temporarily accumulated at the interface). This is shown in Fig.~\ref{fig:left_right_currents} where the left current is represented by continuous lines and the right current by dashed ones. In particular, we show that the correspondence of the two currents is verified both in weak (orange curves) and strong (blue curves) coupling regimes. This is due to the boundness of the local interface operator $V$ in the overall Hamiltonian Eq.~(\ref{eq:overall_Hamiltonian}).

\section{Examples of currents from typical initial conditions}  
\begin{figure}
    \centering
    \includegraphics[width=\columnwidth]{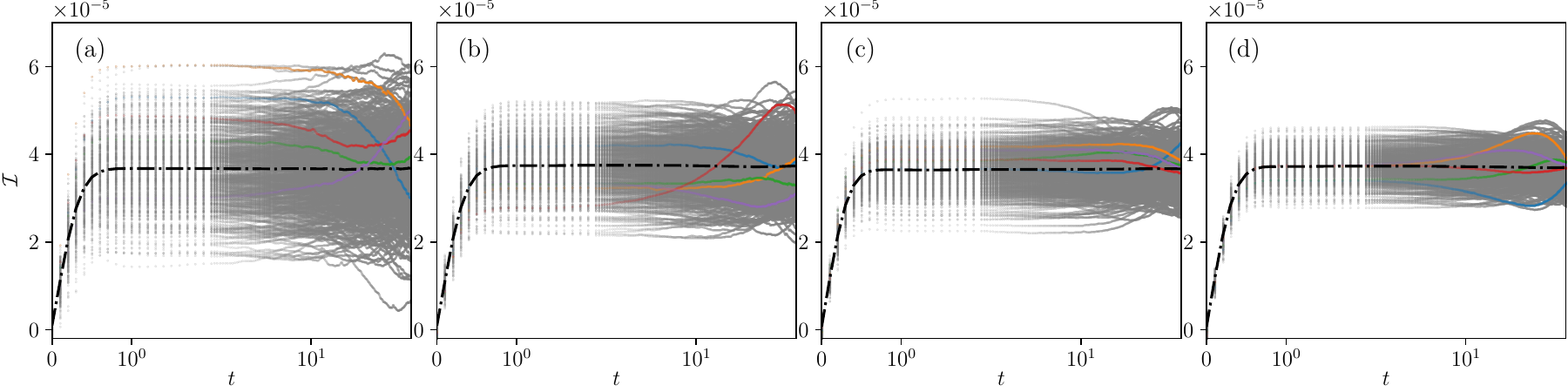}
    \caption{Exact typical currents for different total system sizes (a) $N=22$, (b) $N=24$, (c) $N=26$, and (d) $N=28$ each with 500 typical initial states with a fixed energy per spin $\varepsilon_{\rm L} = -\varepsilon_{\rm R} = 5/12$ and the width of the shell per spin $\delta_{\rm L/R} = 0.1/12 $. The coupling strength $\gamma=0.01$. The averaged current for different system sizes are indicated by the dashed-dotted lines (black). We have randomly selected five typical initial states to demonstrate the current dynamics as indicated by the markers with different colors and the background grey shades includes all the realizations.}     
    \label{fig: sstvsL}
\end{figure}

To give further insight into the dynamics of different typical initial conditions $| \psi_{\rm typ} \rangle$, in Fig.~\ref{fig: sstvsL} we plot $500$ currents versus time, each corresponding to a different typical initial condition (see Eq.~(3) of the main paper). These currents are computed numerically in an exact manner, without perturbative approximations. Of these $500$ initial conditions, we single out a few random ones in different colors to allow the reader to appreciate their dynamics. The black line shows the average of all initial conditions considered. Each panel corresponds to different system sizes, specifically $N=22$, $24$, $26$, $28$ from (a) to (d).        

\section{Details of the optimized time evolution algorithm}

Since the time evolution algorithm used in this work is vital for our large-scale simulations, the details of the numerical implementation are presented here as we believe that they could be useful for future works. 
The evolution for a short time $\delta t$, i.e. $U(\delta t) = \exp(-\im H \delta t)$, can be evaluated using two-site evolution operators $U_{\sigma_n\sigma_m}^{\sigma_n^\prime\sigma_m^\prime}$ after Suzuki-Trotter decomposition \cite{Trotter1959, Suzuki1976}. Here, with $\sigma_n$ we indicate the spin configuration at site $n$, hence a state $|\psi\rangle$ can be represented using the basis state vectors $|\sigma_1, \; \sigma_2, \dots,\sigma_N\rangle$ with the coefficients $c_{\sigma_1, \; \sigma_2, \dots,\sigma_N}$ as 
\begin{align}
|\psi\rangle = \sum_{\sigma_1, \; \sigma_2, \dots,\sigma N} c_{\sigma_1, \; \sigma_2, \dots,\sigma_N}  |\sigma_1, \; \sigma_2, \dots,\sigma_N\rangle. 
\end{align}
Applying the two-site evolution operator $U_{\sigma_n\sigma_m}^{\sigma_n^\prime\sigma_m^\prime}$ to $|\psi\rangle$ thus gives  
\begin{align}\label{eq:gateop}
    c_{\sigma_1\sigma_2\cdots\sigma_i'\cdots\sigma_j'\cdots\sigma_N} = \sum_{\sigma_i, \sigma_j} U_{\sigma_i\sigma_j}^{\sigma_i^\prime\sigma_j^\prime} c_{\sigma_1\sigma_2\cdots\sigma_i\cdots\sigma_j\cdots\sigma_N}. 
\end{align}

\begin{figure}[h!]
    \centering
    \includegraphics[width=0.9\columnwidth]{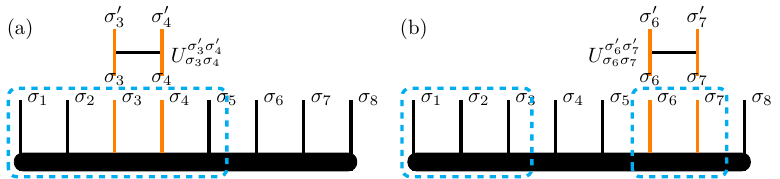}
    \caption{Schemes to aggregate a fixed size matrix for the inner for loop occurred in the two-body gate operation $U$ for case (a) both the two tensor indices are within the first $5$ sites, (b) both tensor indices been contracted are beyond the $5$-th site. The tensor legs inside the blue dashed box mean that they should be placed in the inner loop, while the rest should be placed in the outer loop.}
  \label{fig: gateop}
\end{figure}

One can thus think of $c_{\sigma_1\sigma_2\cdots\sigma_i'\cdots\sigma_j'\cdots\sigma_N}$ as a multi-leg tensor, with each leg taking the possible values $\sigma_i$, as depicted in Fig.~\ref{fig: gateop} .   
At this point one can make two considerations that are the key to significantly speed up the computations:  
1) the memory access complexity and computational complexity are essentially the same (both of the order $O(2^N)$), which is unfriendly to modern computers and 2) the memory access to the quantum state is in general random, which will result in a further slowing down. 
We remark that the memory complexity is here manageable, even on an ordinary laptop, for system sizes of $N=28$ or more because the Suzuki-Trotter algorithm requires the storage of one single vector only. Hence the memory required for $N=28$ is 8 gigabytes for complex double precision. 
More importantly, we use a fused tensor permutation and contraction technique, which is demonstrated in Fig.~\ref{fig: gateop}. Eq. (\ref{eq:gateop}) then means to contract the two tensor indices of the gate operation $U_{\sigma_i\sigma_j}^{\sigma_i^\prime\sigma_j^\prime}$ with the two corresponding indices of the quantum state, which are marked in orange in Fig.~\ref{fig: gateop}. This could be straightforwardly implemented as an outer for loop over the black (uncontracted) legs followed by an inner for loop over the orange (contracted) legs where a small matrix-vector multiplication should be done. 
However, a naive implementation of this approach would in general be very inefficient due to the low computational density and random memory access. The central idea of our technique is to aggregate a fixed number of indices, which is determined by the cache size of the computer. In this case, we choose $5$ indices as a compromise in using the space in the cache and considering the time needed to load and unload the cache. By doing so, the inner loop has an increased computational density because there we implement a multiplication between a $4\times 4$ matrix, which is $U_{\sigma_i\sigma_j}^{\sigma_i^\prime\sigma_j^\prime}$, and $4\times 8$ matrix which represents the values of the coefficients $c_{\sigma_1\sigma_2\cdots\sigma_i'\cdots\sigma_j'\cdots\sigma_N}$ when varying only $5$ indices. 
One further important point, which we depict in Fig.~\ref{fig: gateop}, is that the performance is improved if one considers always sites $1$ to $3$ to be in the inner loop, while the other two sites will depend on where the two-site unitary operator $U_{\sigma_i\sigma_j}^{\sigma_i^\prime\sigma_j^\prime}$ is acting upon. Two examples of the five sites used are enclosed by a dashed blue line in Fig.~\ref{fig: gateop}. 
On top of this, the outer for loop needed to evaluate Eq.~(\ref{eq:gateop}) can be perfectly parallelized since the inner loop only accesses non-overlapping data of the quantum state.

%